\documentstyle[12pt,epsf]{article}

\font\blackboard=msbm10 at 12pt
\font\blackboards=msbm7
\font\blackboardss=msbm5
\newfam\black
\textfont\black=\blackboard
\scriptfont\black=\blackboards
\scriptscriptfont\black=\blackboardss

\newcommand{\junk}[1]{}

\newcommand{\ba}{\begin{array}}
\newcommand{\ea}{\end{array}}
\newcommand{\be}{\begin{equation}}
\newcommand{\ee}{\end{equation}}
\newcommand{\bea}{\begin{eqnarray}}
\newcommand{\eea}{\end{eqnarray}}
\newcommand{\beas}{\begin{eqnarray*}}
\newcommand{\eeas}{\end{eqnarray*}}

\def\identity{{\rlap{1} \hskip 1.6pt \hbox{1}}}

\def\laplace{{\kern1pt\vbox{\hrule height 1.2pt\hbox{\vrule width
1.2pt\hskip
  3pt\vbox{\vskip 6pt}\hskip 3pt\vrule width 0.6pt}\hrule height
  0.6pt}
  \kern1pt}}
\def\scriptlap{{\kern1pt\vbox{\hrule height 0.8pt\hbox{\vrule width
  0.8pt
  \hskip2pt\vbox{\vskip 4pt}\hskip 2pt\vrule width 0.4pt}\hrule height
  0.4pt}
  \kern1pt}}
\def\slash#1{{\rlap{$#1$} \thinspace /}}
\def\roughly#1{\raise.3ex\hbox{$#1$\kern-
.75em\lower1ex\hbox{$\sim$}}}

\def\str{{\rm STr} \,}

\def\tr{{\rm Tr} \,}

\def\r{{\rho}}
\def\s{{\sigma}}
\def\t{{\theta}}
\def\a{{\alpha}}
\def\c{{\chi}}
\def\b{{\beta}}
\def\dr{{\dot{\rho}}}
\def\ds{{\dot{\sigma}}}
\def\da{{\dot{\alpha}}}
\def\db{{\dot{\beta}}}

\def\bX{{\bar{X}}}
\def\bc{{\bar{\chi}}}
\def\bl{{\bar{\lambda}}}
\def\l{{\lambda}}
\def\bt{{\bar{\theta}}}
\def\e{{\epsilon}}
\def\P{{\Phi}}
\def\bP{{\bar{\Phi}}}
\def\z{{\zeta}}
\def\bz{{\bar{\zeta}}}

\newcommand{\gone}[1]{}

\begin{document}
\pagestyle{plain}
\setcounter{page}{1}

\baselineskip16pt

\begin{titlepage}

\begin{flushright}
SU-ITP 01-52\\
hep-th/0112081
\end{flushright}
\vspace{8 mm}

\begin{center}

{\Large \bf Open Dielectric Branes\\}

\end{center}

\vspace{7 mm}

\begin{center}

Mark Van Raamsdonk

\vspace{3mm}
{\small \sl Institute for Theoretical Physics} \\
{\small \sl Stanford University} \\
{\small \sl Stanford, CA, 94306 U.S.A} \\
{\small \tt mav@itp.stanford.edu}\\

\end{center}

\vspace{8 mm}

\begin{abstract}

We derive leading terms in the effective actions describing the
coupling of bulk supergravity fields to systems of arbitrary numbers of 
Dp-branes and D(p+4)-branes in type IIA/IIB string theory. We use these actions 
to investigate the physics of Dp-D(p+4) systems in the presence of weak 
background fields. In particular, we construct various solutions describing 
collections of Dp-branes blown up into open D(p+2)-branes ending on D(p+4)-
branes. The configurations are stabilized by the presence of background fields
and represent an open-brane analogue of the Myers dielectric effect. 
To deduce the D-brane actions, we use supersymmetry to derive
operators corresponding to moments of various conserved currents in 
the Berkooz-Douglas matrix model of M-theory in the presence of
longitudinal M5-branes and then use dualities to relate these
operators to the worldvolume operators appearing in the Dp-D(p+4)-brane 
effective actions.

\end{abstract}

\vspace{0.7cm}
\begin{flushleft}
December 2001
\end{flushleft}
\end{titlepage}
\newpage

\section{Introduction}

One of the most remarkable properties of D-branes in string theory is
that while a single Dp-brane behaves geometrically much like a classical
p-dimensional surface, collections of many Dp-branes can exist in
configurations completely different from those of a set of such classical
surfaces. For example, configurations of $N$ D0-branes
with coordinate matrices that do not mutually commute do not
have well defined positions for the D0-branes \cite{witten}. In some cases, the 
interpretation as a set of $N$ particles is lost entirely, and the
configuration of D0-branes is better described as a smooth, higher 
dimensional object. 

A standard example is the fuzzy sphere \cite{kt1,rey} for which the first 
three coordinate matrices for a set of $N$ D0-branes are identified with the 
generators of the $N$-dimensional irreducible representation of
$SU(2)$, $X^i = \alpha J^i$. This configuration, for which  
\be
\label{fuzzy}
[X^i, X^j] = \alpha i \epsilon^{ijk} X^k \; ,
\ee
has an alternate description as a spherical D2-brane of radius $R \approx 
{\alpha N \over 2}$ with a 
uniform magnetic field on its worldvolume (corresponding to the zero-brane 
charge). A second example, with an infinite number of D0-branes, is given by 
choosing $X^i$ such that
\[
[X^i, X^j] = i \theta^{ij} \; 
\]
for constant $\theta$ \cite{bss}.
In this configuration, the coordinate matrices represent the algebra of 
noncommutative $R^{2n}$ and the physical interpretation is an infinite flat 
D(2n)-brane with a uniform magnetic field $F_{12} = \dots = F_{(2n-1) \; (2n)}$.
This ability to construct higher dimensional brane configurations using 
D0-branes 
is essential for the success of the BFSS Matrix Theory \cite{bfss},
where for example 
arbitrary membrane configurations in M-theory must be described in terms of 
the low energy degrees of freedom of D0-branes. 

The examples above indicate that while noncommuting configurations do not 
have well defined positions for the individual D0-branes, there is still some 
geometrical interpretation. In fact, it is still possible to measure the 
spacetime distribution of matter and charges for an arbitrary noncommuting 
configuration using the operators coupling to bulk supergravity fields in the 
low energy effective action. For example, the zero-brane charge distribution is 
measured by the operator coupling to the time component of the Ramond-Ramond 
one-form field, and is given (at weak coupling and small $\alpha'$) in momentum 
space by \cite{tv0}
\be
\label{j0}
{\cal J}_{(1)}^0(k) = \tr(e^{i k \cdot X}) \; .
\ee
Operators measuring multipole moments of the zero-brane charge distribution 
correspond to derivatives with respect to $k$ of this expression at $k=0$. 
Similarly, the D(2p+1)-brane charge density is measured by operators\footnote{
Throughout this paper, $\str$ will denote a 
symmetrized trace in which one averages over all orderings of factors in the 
trace, with commutators treated as a unit in the symmetrization. In
particular, the individual terms in the expansion of the exponential
should be symmetrized with the remaining factors in the trace.}
\be
\label{jn}
{\cal J}_{(2n+1)}^{0 i_1 \cdots i_{2n}}(k) \propto \str([X^{[i_1}, X^{i_2}] 
\cdots [X^{i_{2n-1}}, 
X^{i_{2n}]}] e^{i k \cdot X})
\ee
coupling to the higher Ramond-Ramond fields in the effective 
action. From this 
expression it is manifest that noncommuting configurations of D0-branes involve 
higher dimensional brane charges. These operators and the corresponding 
bulk-brane effective actions (for weak coupling and small $\alpha'$) were worked
out 
for D0-branes in \cite{tv0} and for Dp-branes of any dimension in 
\cite{tvp,myers}.

Of course, these effective actions also permit the study of collections of 
D-branes in the presence of weak background fields, and in particular provide a 
method to produce stable noncommutative brane configurations with higher 
dimensional brane charges.  For example, Myers showed that in the presence of a 
constant Ramond-Ramond three-form field strength, a system of D0-branes will 
blow up into the spherical D2-brane configuration (\ref{fuzzy}) with a radius 
proportional to the field strength \cite{myers}. 

Thus, the low energy effective actions describing linear couplings of bulk 
supergravity fields to the D-brane worldvolume fields provide essential tools 
in understanding the spacetime properties of general noncommuting D-brane 
configurations. \\
\\
{\bf Open branes}\\
\\
So far, we have discussed collections of a single type of D-brane in the bulk, 
and the higher dimensional branes that arose from noncommuting configurations 
were ``closed'' branes in the sense that they have the geometry of surfaces 
without boundaries, and carry zero net higher dimensional brane charges in the 
case of finite $N$ (due to the vanishing of the expressions (\ref{jn}) at 
$k=0$). On the other hand, it is well known that branes can often have 
boundaries on higher dimensional branes \cite{strominger}. Perhaps the simplest 
example is that a Dp-brane can end on D(p+2)-brane (related by S and T-duality 
to the fact that fundamental strings may end on D-branes). Given that D-branes 
can exist in higher dimensional closed brane configurations, it is natural to 
ask whether we can also build open brane configurations from lower dimensional 
branes. For example, can we find configurations of Dp-branes corresponding to 
open D(p+2)-branes ending on D(p+4)-branes? 

A hint that the answer is yes is provided by the matrix model of Berkooz and 
Douglas, proposed to describe M-theory in the presence of longitudinal 
M5-branes \cite{bd} (see also \cite{li}). If correct, this model should
include states corresponding to 
arbitrarily shaped open membranes ending on the M5-branes, since these open 
membranes are allowed objects in M-theory. On the other hand, the degrees of 
freedom of the model are the lowest energy modes of 0-0 strings and 0-4 strings 
in a system of $N$ D0-branes and $k$ D4-branes, where $N$ represents the DLCQ 
momentum in the theory and $k$ corresponds to the number of $M5$-branes. Thus, 
we should be able to describe arbitrary open membranes in terms of the degrees 
of freedom of D0-branes in the presence of D4-branes. Interpreted in the context 
of type IIA string theory at low energies and weak coupling, these 
configurations should correspond to D0-branes blown up into open D2-branes 
ending on the D4-branes.

In this paper, motivated by the Berkooz-Douglas matrix model, we will consider 
systems of Dp-branes in the presence of D(p+4)-branes and look for classical 
configurations in which the Dp-branes form a noncommutative open D(p+2)-brane 
ending on the D(p+4)-branes. Just as for the case of noncommutative closed-brane 
configurations, an essential tool will be the low energy effective action 
describing the couplings of the brane system to the bulk supergravity fields. 
Since this effective action has not to our knowledge been determined previously 
for the Dp-D(p+4) system with arbitrary number of Dp-branes and D(p+4) branes, 
our first task will be to provide a derivation of some of the leading terms in 
this action. 

Thus, the goals of this paper will be to derive leading terms in the effective 
action describing the couplings of the Dp-D(p+4) system to bulk supergravity 
fields and to use this effective action to create and study open noncommutative 
branes. 

A summary of the remainder of the paper is as follows.
In section 2, we discuss various tools that will be useful in constructing the 
Dp-D(p+4) effective actions. We recall that all the actions are related by 
T-duality and further that these actions are related to the action for Matrix 
theory (in this case, the Berkooz-Douglas matrix model) in the presence of 
background 11-dimensional supergravity fields. We will find it simplest to 
derive the currents in the matrix model. In section 3, we show that all the 
currents are related to each other by supersymmetry and that in the matrix 
model context, these supersymmetry relations may be used to derive all the 
currents from a single ``primary'' current, namely the operator $T^{++}$ 
coupling to the metric component $h_{++}$ (corresponding in the D0-D4 picture to
the zero-brane density $J_0$). In section 4, as a check of our approach, we use 
these supersymmetry relations in the context of the BFSS matrix model to 
rederive the currents for systems of a single type of D-brane starting from the
expression (\ref{j0}). We find complete agreement with known
expressions for the matrix theory currents derived in \cite{kt2, tvm,
dnp}. 

In section 5, we recall the Berkooz-Douglas matrix model, obtained from the 
dimensional reduction of the D5-D9 action in flat space. In section 6, we use 
our supersymmetry relations to derive leading terms in the operators describing 
conserved currents in the Berkooz-Douglas model. In this case, we do not even 
know the ``primary'' operator $T^{++}$ except that it should agree with the BFSS 
result (\ref{j0}) when the fields arising from 0-4 strings are set to zero. 
Nevertheless, we are able to deduce the leading new terms (quadratic in the 0-4 
strings) by demanding consistency of the supersymmetry relations. 
This is enough to determine leading operators in the Dp-D(p+4) actions
coupling to arbitrary weak type II supergravity fields, and we present these
results in section 7. 

In section 8, we apply our results to look for noncommutative open-brane 
configurations. We first discuss configurations of D0-branes corresponding to 
flat open D2-branes of various shapes inside a D4-brane. In particular, we show 
explicitly that planar configurations of noncommutative instantons carry 
non-zero D2-brane area, as suggested previously by Berkooz in the context of the 
matrix model for the noncommutative (0,2) theory \cite{berkooz}. Since these 
instanton 
configurations are supersymmetric, we may conclude that the flat noncommutative 
D2-brane solutions we discuss are stable and BPS. Starting from a disk-like open 
D2-brane configurations, we show that by turning on a gradient of either the RR 
one-form or the RR three-form potential in a direction perpendicular
to the D4-brane, we can pull the interior of the open D2-brane off the
D4-brane such that 
the final configuration is a bulging parabolic D2-brane with a circular boundary 
on the D4-brane (for a preview, see figure 6 in section 8.4). Thus, starting with
a collection of coincident D0-branes, we 
can turn on a combination of background fields to produce an ``open dielectric 
brane'' analogous to closed dielectric brane discovered by Myers. 

In section 9, we discuss an approach based on the ADHM construction for deriving 
higher order terms in the 
expression for the primary current $T^{++}$ in the Berkooz-Douglas model that 
would allow a more complete derivation of the Dp-D(p+4) effective actions.   
Finally, we offer some concluding remarks in section 10 and a variety of useful 
formulae and results in a set of appendices.

\section{Deriving D-brane actions}

We would like to derive leading terms in the effective actions describing the 
couplings of type IIA/IIB supergravity fields to the worldvolume fields of a 
system of Dp-branes and D(p+4)-branes. These worldvolume fields arise from the 
massless 
excitations of open p-p strings, p-(p+4) strings, and (p+4)-(p+4) strings. The 
(p+4)-(p+4) fields propagate on the (p+5)-dimensional worldvolume of the 
D(p+4)-branes, while the p-p and p-(p+4) fields are restricted to the 
(p+1)-dimensional worldvolume of the Dp-branes. The field content and flat space 
Lagrangian of the theory will be reviewed in section 5.

In general, the effective action for the couplings of type II supergravity 
fields to the worldvolume fields on a system of D-branes takes the form
\be
\label{sugyact}
S = \int d^{10} x (  {1 \over 2} h_{\mu \nu} {\cal T}^{\mu \nu} + \phi {\cal 
J}_\phi + {1 \over 2} B_{\mu \nu} {\cal J}_s^{\mu \nu}+ {1 \over n!} 
C^{(n)}_{\mu_1 \cdots \mu_n} {\cal J}_{(n)}^{\mu_1 \cdots \mu_n})  
\ee
Here $h$ is the metric fluctuation, $\phi$ is the dilaton, $B$ is the NS-NS two 
form field, and $C^{(n)}$ are the Ramond-Ramond fields, with $n$ even in the 
type IIB case and odd in the type IIA case. Our goal is to determine expressions 
for the currents ${\cal T}$, ${\cal J}_\phi$, ${\cal J}_s$, and ${\cal J}_{(n)}$ 
in terms of the D-brane 
worldvolume fields. In principle, one could compute these directly 
by calculating tree-level string amplitudes with one closed string vertex 
operator and various numbers of open string vertex operators on the disk but 
this would be a forbidding amount of work. Fortunately, there are a number of 
indirect approaches which help to determine these actions, and we review some of 
these presently.\\
\\
{\bf Symmetries and T-duality}\\
\\
Firstly, the various symmetries of the theory place strong constraints on the 
possible terms appearing in the action. For the Dp-D(p+4) system, we must have 
$SO(p,1)$ Lorentz invariance in the Dp-brane directions, $SO(4)$ rotational 
invariance in the D(p+4) directions transverse to the Dp-brane, and $SO(5-p)$ 
rotational invariance in the directions transverse to the D(p+4)-branes. 

Furthermore, the actions are all related by T-duality, which acts on the 
worldvolume fields by dimensional reduction/oxidation, and acts on the bulk 
fields (to linear order) as \cite{bho}
\bea
h_{\mu \nu} &\rightarrow& h_{\mu \nu} \nonumber\\
B_{\mu \nu} &\rightarrow& B_{\mu \nu} \nonumber \\
h_{\mu \hat{\nu}} &\leftrightarrow& - B_{\mu \hat{\nu}} \nonumber \\
h_{\hat{\mu} \hat{\nu}} &\rightarrow& -h_{\hat{\mu} \hat{\nu}} 
\label{tduality}\\
B_{\hat{\mu} \hat{\nu}} &\rightarrow& -B_{\hat{\mu} \hat{\nu}} \nonumber \\
\phi &\rightarrow& \phi - {1 \over 2} \sum_{\hat{\mu}} h_{\hat{\mu} \hat{\mu}} 
\nonumber \\
C^{(n)}_{\mu_1 \cdots \mu_{n-k}} {}^{\hat{\nu}_1 \cdots \hat{\nu}_k} 
&\rightarrow& {1 
\over (n-k)!} \e^{\hat{\nu}_1 \cdots \hat{\nu}_m} C^{(n-2k+m)}_{\mu_1 \cdots 
\mu_{n-k} \hat{\nu}_{k+1} \hat{\nu}_m} \nonumber 
\eea
where hatted indices are in the $m$ directions being dualized and unhatted 
indices denote the remaining directions. Thus, for example the operator coupling 
to $h_{08}$ in the D0-D4 action can be obtained from the operator coupling to 
$-B_{08}$ in the D5-D9 action by dimensional reduction. These T-duality 
relationships proved particularly useful in the construction of the nonabelian 
actions for a system of Dp-branes, where T-duality combined with knowledge of 
the abelian D9-brane action determines much of the nonabelian structure in the 
dual Dp-brane actions {\cite{tvp, myers, tseytlin}}.\\  
\\
{\bf Conservation Relations}\\
\\
Most of the supergravity fields we are interested in are gauge fields
and transform nontrivially under gauge transformations such as
\[
\delta h_{\mu \nu} = \partial_{(\mu} \xi_{\nu)}
\] 
In order for the action (\ref{sugyact}) to be gauge invariant, the currents must
obey conservation laws, which to linear order take the form
\[
\partial_\mu I^{\mu \nu_1 \cdots \nu_n} = 0
\]
In cases where the expressions for the currents are unknown, these
relations will help to determine certain components of the currents
in terms of other components.\\ 
\\
{\bf Relation to Matrix Theory}\\
\\
Another tool that has proved very useful in constructing non-abelian Dp-brane 
actions is the relationship between Matrix theory and D0-branes in type IIA 
string theory. The usual BFSS Matrix Theory lagrangian arises from leading terms 
in the low energy, weak coupling action for D0-branes in flat space. In a 
similar way, a matrix model action describing M-theory in the presence of weak 
eleven-dimensional supergravity fields arises from leading terms in the action 
describing a system of D0-branes in the presence of weak type IIA supergravity 
fields \cite{tv0}. Using this relationship, it is therefore possible to derive 
leading terms in the D0-brane action (and by T-duality, the other Dp-brane 
actions) from the action for Matrix Theory with background fields, as was 
carried out in \cite{tv0}. Explicitly, the spacetime currents (\ref{sugyact}) 
for a system of D0-branes are determined in terms of the Matrix theory currents 
as\footnote{Here, ${\cal O} 
(X^n)$ indicates terms for which the number of bosonic fields plus the number of 
derivatives plus ${3 \over 2}$ the number of fermionic fields is $n$ or more, 
i.e. $n$ is the mass dimension in the usual four dimensional counting.}
\begin{equation}
\begin{array}{ll}
{\cal T}^{00}  = T^{++} +  T^{+-} + {\cal O} (X^{8})
&{\cal J}_{(1)}^0 = T^{++} \\
{\cal T}^{0i} = T^{+i} + T^{-i} + {\cal O} (X^{10})
&{\cal J}_{(1)}^i = T^{+i} \\
{\cal T}^{ij} = T^{ij} + {\cal O} (X^{8})
&{\cal J}_{(3)}^{ijk}  =  J^{ijk} + {\cal O} (X^{8})  \\
{\cal J}_\phi = T^{++} -  \left({1 \over 3} T^{+-}  + {1
\over 3} T^{ii} \right) + {\cal O} (X^{8})  &
{\cal J}_{(3)}^{0ij}  =   J^{+ ij}  + {\cal O} (X^{10})
\\
{\cal J}_s^{0i}  =  {1 \over 2} J^{+ -i} +{\cal O} (X^8)&
{\cal J}_{(5)}^{0ijkl}  =   M^{+ -i jkl}+{\cal O} (X^8) \\
{\cal J}_s^{ij}  =  {1 \over 2} J^{+ ij}  - {1 \over 2} J^{-ij} + {\cal O} 
(X^{10})
&{\cal J}_{(5)}^{ijklm}  =  - M^{-ij klm} +{\cal O} (X^{10})
\end{array}
\label{relationship}
\end{equation}
where $T$, $J$, and $M$ are the Matrix theory stress-energy tensor, membrane 
current and fivebrane currents which couple to the eleven-dimensional 
supergravity fields as
\be 
\label{mtact}
{\cal L}_{MT} = {1 \over 2} h_{IJ} T^{IJ} + { 1 \over 3!} A_{IJK} J^{IJK} + { 1 
\over 6!} A^D_{IJKLMN} M^{IJKLMN} + i S^{I} \psi_I \; .
\ee
Here, $h$ is the metric, $A$ is the three-form field, and $A_D$ is the six-form 
field with field strength dual to the field strength of $A$. For future use, we 
also include a fermionic current $S^I$ which couples to the gravitino $\psi_I$. 
It is important to note that here and in the rest of this work, $T$, $J$, $M$ and $S$ represent matrix theory currents integrated over the longitudinal direction 
so that the resulting expressions depend only on time and the nine transverse 
directions.

While the results above were derived for a system of D0-branes and the BFSS 
Matrix model, an identical relationship should exist between the
action for the D0-D4 system and the Berkooz-Douglas matrix model for
M-theory in the presence of 
M5-branes. Precisely the same limit relates this matrix model to the D0-D4 
system as relates the BFSS model to the system of D0-branes. Thus, the relations 
(\ref{relationship}) should determine leading terms in the currents
for the D0-D4 system if the 
currents on the right side are taken to be those of the Berkooz-Douglas 
model.

In this paper, we will find it most convenient to derive the currents in the 
Berkooz-Douglas 
model, then use the relations (\ref{relationship}) to determine leading 
terms in the currents for the D0-D4 system, and finally, determine the effective 
actions for the general Dp-D(p+4) systems via T-duality.

For the BFSS model, the currents were determined in \cite{kt2,tvm} by comparing 
the one-loop matrix theory effective action for a pair of arbitrary widely 
separated systems with the classical effective potential obtained from 
linearized DLCQ eleven-dimensional supergravity.\footnote{Recently, the bosonic 
parts of these currents have also been computed directly from string theory 
\cite{oo1,oo2}.} One way to determine the 
currents in the Berkooz-Douglas model would be to perform a similar one loop 
matrix theory 
calculation. In this case, there is only half as much supersymmetry, so the 
terms that could be reliably compared with supergravity are at lower orders  
($F^2/r^3$ type terms rather than $F^4/r^7$ terms). 

While the calculation of the Berkooz-Douglas matrix model potential seems 
feasible,
we will take an approach that is still less direct and constrain the
currents using supersymmetry.\\
\\
{\bf Supersymmetry}\\
\\
We have argued above that various bosonic symmetries place strong constraints on 
the terms in the effective action. Further constraints follow from the fact that 
the effective actions in string theory should be supersymmetric. 

To understand the constraints that supersymmetry places on the currents, 
consider a general supergravity theory with bosonic fields denoted by $\phi$ and 
fermionic fields denoted by $\psi$ coupled to a system of branes. The linear 
couplings between supergravity fields and the worldvolume fields on the branes 
will take the form
\[
S = \int d^dx \left\{ \phi J + \bar{\psi} S \right\}
\]
where $J$ represents the set of (unknown) bosonic currents and $S$ represents 
the (unknown) fermionic currents. This effective action should be invariant 
under the supersymmetries of the theory, so 
\be
0 = \delta S = \int d^dx \left\{ \delta \phi J + \bar{\psi} \delta S 
+ \phi \delta J + \delta \bar{\psi} S \right\}
\label{suvar}
\ee
Here, for example, $\delta J$ is the variation of the bosonic current under a 
supersymmetry transformation of the worldvolume fields. To linear order, the 
supersymmetry variations of the bulk supergravity fields will be some
known expressions given schematically 
by
\[
\delta \phi = \bar{\psi} \gamma \epsilon \qquad \delta \psi = \e \partial \phi
\]
where $\e$ is the supersymmetry variation parameter, $\gamma$ is some product of 
Dirac matrices, and $\partial$ is some operator containing a single derivative.

Inserting these bulk supersymmetry variations into the expression (\ref{suvar}), 
we obtain
\[
\int d^dx \left( \bar{\psi} \{ \delta S + \gamma \e J \} + \phi \{ \delta J - 
\bar{\e} \partial S \} \right) = 0
\]
where we have integrated by parts to obtain the final term.

Naively, it would seem that we could now set each of the expressions in curly 
brackets to zero to obtain one relation for each component of each bulk 
supergravity field. However, we must remember that the supersymmetry variation 
of the action is only required to vanish after using the bulk equations of 
motion (if we are using an on-shell formulation of supergravity). Writing these 
equations of motion as 
\[
\slash{D} \psi = 0 \qquad D^2 \phi = 0
\]
the proper conclusion is that the worldvolume currents obey the relations
\bea
\delta S &=& -\gamma \e J + \slash{D} R_{ferm} \nonumber\\
\delta J &=& \bar{\e} \partial S + D^2 R_{bos}
\label{susyrel}
\eea
where $R_{bos}$ and $R_{ferm}$ are some ``auxiliary'' currents. Thus, invariance 
of the effective action under supersymmetry implies a set of equations relating 
the supersymmetry variation of bosonic currents to fermionic currents and 
vice-versa. We will find that these relations are very useful in actually 
deriving expressions for the various worldvolume currents based on knowledge of 
a single current.

The approach just described is quite general and for our purposes could be used 
either in the case of type II string theory directly to obtain relations between 
currents in the Dp-D(p+4) system, or in the context of Matrix Theory, to obtain 
relations between the Matrix theory currents. In this paper, we will take the 
latter approach, since as we will explain shortly, the form of the relations 
(\ref{susyrel}) in the matrix theory case are particularly useful for deriving 
the currents. 

A very similar approach was used to derive expressions for vertex operators
in string theory in \cite{gs} and more recently to derive 
vertex operators for the eleven-dimensional superparticle in \cite{ggk} and for 
the eleven-dimensional supermembrane in \cite{dnp}. Our approach is slightly 
different in that we are working with an off-shell
effective action rather than the operators corresponding to particular
on-shell states.   

\section{Supergravity couplings in matrix theory via supersymmetry}

In this section, we flesh out the general procedure just described to derive 
explicit supersymmetry relations between the currents in Matrix
Theory. A related discussion for the continuum supermembrane may be
found in \cite{dnp, plefka}. In the present 
case, the bulk fields are those of eleven-dimensional supergravity. We use 
conventions for which the kinetic terms in the eleven-dimensional supergravity 
lagrangian are
\[
{\cal L}_{\rm kin} = {1 \over 2 \kappa^2} \left \{ \sqrt{-g} R - {1 \over 48} 
F_{IJKL} F^{IJKL} - 2i \bar{\psi}_I \Gamma^{IJK} \partial_J \psi_K \right\} \; .
\]
To linear order, the supersymmetry variations of the fields in this Lagrangian 
are given by
\bea
\delta h_{IJ} &=& 2i \bar{\epsilon} \Gamma_{(I} \psi_{J)} \nonumber\\
\delta A_{IJK} &=& 3i \bar{\epsilon} \Gamma_{[IJ} \psi_{K]} \label{susy1}\\
\delta \psi_I &=& - {1 \over 2} \partial_J h_{KI} \Gamma^{JK} \epsilon - {1 
\over 72} (\Gamma_I {}^{JKLM} - 8 \delta_I^J \Gamma^{KLM}) \e \partial_{[J} 
A_{KLM]} \nonumber
\eea
These supergravity fields couple linearly to the matrix theory
currents as
\be 
\label{mtact2}
{\cal L}_{MT} = {1 \over 2} h_{IJ} T^{IJ} + { 1 \over 3!} A_{IJK} J^{IJK} + { 1 
\over 6!} A^D_{IJKLMN} M^{IJKLMN} + i S^{I} \psi_I \; .
\ee
In order to write the supersymmetry variation of this expression, we need to 
know how the dual six-form field varies under a supersymmetry transformation. It 
may 
be checked that the variation
\be
\label{susy2}
\delta A^D_{IJKLMN} = 6i \bar{\e} \Gamma_{[IJKLM} \psi_{N]}
\ee
leads to the correct supersymmetry variation for the seven-form field strength 
(consistent with the variation of the dual four-form field strength). 

Inserting the bulk supersymmetry transformations (\ref{susy1}),(\ref{susy2}) 
into 
the 
action (\ref{mtact2}), and separating terms involving the metric, gravitino and 
form-fields, we find
\beas
0 &=& \int d^{11} x \left\{ h_{IJ} \delta T^{IJ} - i \partial_I h_{JK} S^K 
\Gamma^{IJ} \e \right\} \\
0 &=& \int d^{11} x \left\{ i \delta S^I \psi_I + i \bar{\e} \Gamma_I \psi_J 
T^{IJ} + {i \over 2} \bar{\e} \Gamma_{IJ} \psi_K J^{IJK} + {i \over 120} 
\bar{\e} \Gamma_{IJKLM} \psi_N M^{IJKLMN} \right\}\\
0 &=& \int d^{11} x \left\{ A_{IJK} \delta J^{IJK} + {1 \over 120} A^D_{IJKLMN} 
\delta M^{IJKLMN} - {i \over 48} F_{IJKL} S^M (\Gamma_M {}^{IJKL} - 8 
\delta_M^I \Gamma^{JKL}) \e \right\}
\eeas
In order for the last equation to hold, the supersymmetry variations of $J$ and 
$M$ should take the form
\beas
\delta J^{IJK} &=& \partial_L K^{LIJK}\\
\delta M^{IJKLMN} &=& \partial_P N^{PIJKLMN}
\eeas
where $K$ and $N$ are totally antisymmetric, so that we may integrate by parts 
to obtain an expression depending only on the field strength.

To determine the appropriate relations (\ref{susyrel}), we need to take into 
account the equations of motion for the bulk fields, which to linear order read
\beas
0 &=& \partial_I \partial_K h^K{}_J + \partial_J \partial_K h^K {}_I - 
\partial^2 h_{IJ}  - \partial_I \partial_J h^K {}_K\\
0 &=& \partial^I F_{IJKL}\\
0 &=& \Gamma^{IJK} \partial_J \psi_K
\eeas
Then the desired supersymmetry relations between the currents are
\bea
\delta T^{IJ} &=& -i \partial_K S^J \Gamma^{KI} \e + \{\partial_K \partial^J 
R^{KI} + \partial_K \partial^I 
R^{KJ} - \partial^2 R^{IJ} - \partial_K \partial_L R^{KL} \eta^{IJ} \} 
\label{master}\\
\delta S^I &=& - \Gamma^0 \Gamma_J \e  T^{JI} - {1 \over 2} \Gamma^0 \Gamma_{JK} 
\e
J^{JKI} - {1 \over 120} \Gamma^0 \Gamma_{JKLMN} \e M^{JKLMNI} - \{ \Gamma^0 
\Gamma^{IJK} \partial_J R_K \}  \nonumber
\eea
\[
K^{IJKL} + {1 \over 5040} \e^{IJKL} {}_{M_1 \cdots M_7} N^{M_1 \cdots M_7} =  
- { i \over 12} S^M (\Gamma_M {}^{IJKL} - 8 \delta_M^I \Gamma^{JKL}) \e + 
\{ \partial^{[I} R^{JKL]} \}
\]
where the terms in curly brackets are auxiliary terms as in (\ref{susyrel}) 
above. 
An important property of the matrix theory currents is that they have dimensions 
determined by their $SO(1,1)$ charge $q$. In the limit defining Matrix Theory 
from type IIA string theory, the only currents which survive are those obeying
\[
d_{bos} = 4 - 2q
\]
for bosonic currents and
\[
d_{ferm} = {9 \over 2} - 2q 
\]
for fermionic currents \cite{tv0}. In these expressions, $d$ is the dimension in
units 
where bosonic fields and time derivatives are assigned dimension 1, fermionic 
fields are assigned dimension 3/2, and transverse momenta are assigned dimension 
-1. 

For the bosonic currents, $q$ is simply the number of $+$ indices minus the 
number of $-$ indices. Thus, there is a unique current $T^{++}$ of lowest 
dimension 0 and a unique current $T^{--}$ of highest dimension 8. For the 
fermionic currents $S^I$, $q$ is the number of $+$ indices minus the number of 
$-$ indices plus the eigenvalue of ${1 \over 2} \Gamma^{-+}$. In particular, the 
32 supersymmetry generators (zero momentum part of the currents coupling to the 
time component of the gravitino field) split into 16 with dimension ${3 \over 
2}$, denoted by $S_+^+$, and 16 with dimension $7/2$, denoted by $S_-^+$ (where 
the lower sign denotes the $\Gamma^{-+}$ eigenvalue). In terms of these 
generators the supersymmetry variations of a given operator are 
\[
\delta_+ {\cal O} = [\e^\dagger S^+_+, {\cal O}] \; , \qquad \qquad 
\delta_- {\cal O} = [\e^\dagger S^+_-, {\cal O}]
\]
Comparing these with (\ref{master}), it is easy to see that the $S^+_+$ will act 
as lowering operators, giving lower dimensional currents in terms of higher 
dimensional currents, while the $S^+_-$ will act as raising operators, giving 
higher dimensional currents in terms of lower dimensional currents. 

It is then very useful to split up the relations (\ref{master}) into independent 
equations with specific $SO(1,1) \times SO(9)$ transformation
properties. Since the Matrix theory variables depend only on time, it
is convenient in practice to write the matrix theory currents in
(spatial) momentum space, as functions of time and nine transverse momenta.
As noted previously, we will always talk about currents integrated
over the longitudinal ($x^-$) direction so the relations we use will
be the spatial Fourier transforms of the expressions (\ref{master}) at
longitudinal momentum $k_-$ equal to zero.

For example, the raising supersymmetries acting on the stress energy tensor 
give\footnote{To avoid unwieldy expressions, we do not explicitly write the 
terms arising from the auxiliary currents.} 
\beas
\delta_- T^{++} &=& - k_i S^+_+ \Gamma^{i+} \epsilon_- + \{R^{IJ} {\rm 
terms}\}\\
\delta_- T^{+i} &=& - {i \over 2} \partial_t S^+_+ \Gamma^{+i} \e_- 
- { 1\over 2}  k_j S_+^i \Gamma^{j+} \e_- - { 1\over 2} k_j S_-^+
\Gamma^{ji} \e_- + \{R^{IJ} {\rm terms}\}\\
\delta_- T^{+-} &=& - {1 \over 2} k_i S^-_+ \Gamma^{i+} \e_- + \{R^{IJ} {\rm 
terms}\}\\
\delta_- T^{ij} &=& -i \partial_t S^i_+ \Gamma^{+j} \e_- - k_k S^i_- 
\Gamma^{kj} \e_- + \{R^{IJ} {\rm terms}\}\\
\delta_- T^{-i} &=& -{i \over 2} \partial_t S^-_+ \Gamma^{+i} \e_- - {1 \over 
2} 
k_j S^-_- \Gamma^{ji} \e_- + \{R^{IJ} {\rm terms}\}\\
\delta_- T^{--} &=& 0
\eeas     
Here, all currents are written as functions of time and transverse spatial 
momenta, and $\partial_t$ is the derivative with respect to $x^+$
which is identified with worldvolume time.

If not for the auxiliary current terms, these relations (and the others we have 
not written explicitly) would completely determine all currents from the lowest 
dimension current $T^{++}$ which measures the density of D0-brane charge. In 
fact, we will find that the auxiliary currents do not introduce much ambiguity 
and are often completely determined as the only possible expressions that can be 
subtracted from the variations of the currents on the left hand side of 
(\ref{master}) to give an expression of the correct form to match the right hand 
side. 

Roughly speaking, the set of currents $T$, $J$, $M$, and $S$ form a multiplet 
under the supersymmetry with $T^{++}$ playing the role of a primary field. The 
complete set of supersymmetry relations are summarized pictorially in figure 1, 
though one should keep in mind the presence of the
auxiliary terms and also terms involving time derivatives of
currents.\footnote{The currents $M^{ijklmn}$ and $M^{+ijklm}$
correspond to transverse fivebranes and seem to be identically zero in
Matrix Theory. In addition to the currents mentioned, there is
also a current which gives rise to the D6-brane current in type IIA
string theory \cite{tvm}. This may come in to the supersymmetry relations,
however in this paper we ignore it since we will mostly be interested
in currents with dimension less than or equal to 4, while the sixbrane current 
has components with dimensions 6 and 8.}
\begin{figure}
\centerline{\epsfysize=4truein \epsfbox{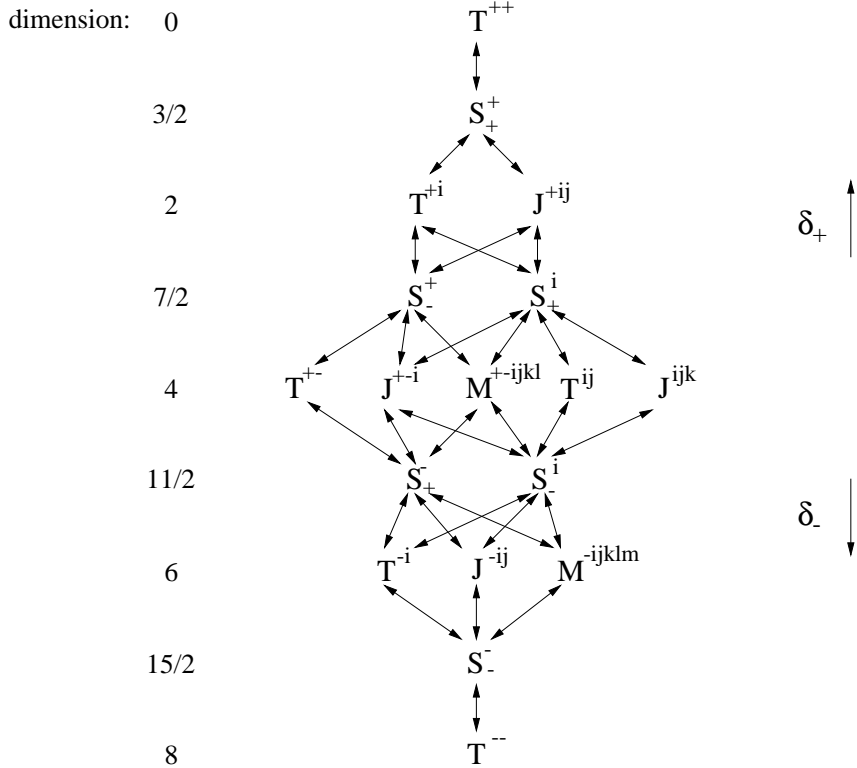}}
 \caption{Summary of SUSY relations between Matrix theory currents.}
\end{figure}

In the next section, we will test our approach by checking that the
supersymmetry relations are satisfied on the known expressions for
the currents in the BFSS theory. We will then apply the technique to
derive currents in the Berkooz-Douglas model in section 6 after
reviewing the model in section 5.
 
\section{Currents in the BFSS matrix model}

In this section, as a check of our approach, we use the supersymmetry
relations to rederive currents in the BFSS Matrix model starting with
the lowest dimension current $T^{++}$. We will find complete agreement
with the known results, derived originally in \cite{kt2,tvm,dnp,plefka}.

The Lagrangian for the theory corresponding 
to $N$ units of longitudinal momentum is given by the low energy theory of 
N D0-branes in flat space, namely the dimensional reduction of ten-dimensional 
$U(N)$ SYM theory to 0+1 dimensions,
\be
\label{bfss}
{\cal L} = \tr( -{1 \over 2} D_0 X^i D_0 X^i + {1 \over 4} [X^i,
X^j]^2 - {i \over 2} 
\bl \gamma^0 D_0 \l + {1 \over 2} \bl \gamma^i [X^i, \l] )
\ee
where $X^i$, $i=1, \dots, 9$ are scalars in the adjoint of $U(N)$ and $\lambda$ 
are 32-component Majorana-Weyl spinors. 

The supersymmetry transformation rules include the 16 supersymmetries inherited 
from those of the $D=10$ SYM theory,\footnote{Often we will write 
ten-dimensional expressions to denote their dimensionally reduced counterparts, 
for example, $F_{ij} \equiv i[X^i, X^j]$.}  
\bea
\partial_- \lambda &=& {1 \over 2} F_{ij} \gamma^{ij} \e_- + F_{0i} \gamma^{0i} 
\e_-
\nonumber \\
\partial_- X_i &=& -i \bar{\e}_- \gamma_i \lambda \label{raising}
\eea
as well as 16 linearly realized supersymmetries (which act non-trivially on a 
D0-brane to generate the 256 polarization states)
\bea
\partial_+ \lambda &=& \gamma_0 \e_+ \identity \nonumber\\
\partial_+ X^i &=& 0 \label{lowering}
\eea
Here, $\e_+$ and $\e_-$ are positive and negative chirality spinors, satisfying
\[
{1 \over 2}(1 \pm \gamma^{10})\e_\pm = \e_\pm
\]
From the discussion of the previous section, it is clear that the first set 
corresponds to the ``raising'' supersymmetries generated by $S^+_-$ while the 
second set corresponds to the ``lowering'' supersymmetries generated by $S^+_+$. 

In order to apply the supersymmetry relations (\ref{master}) explicitly, we must 
match conventions between those in (\ref{master}) with those in (\ref{bfss}), 
(\ref{raising}), and (\ref{lowering}). In particular, we have   
\be
\label{gtrans}
\Gamma^- = {1 \over \sqrt{2}} (1 - \gamma^{10}) \gamma^0 \qquad \Gamma^+ = {1 
\over \sqrt{2}} (1+ \gamma^{10}) \gamma^0 \qquad \Gamma^i = \gamma^{i0} 
\ee
where the upper-case Dirac matrices are those appearing in the 
eleven-dimensional supergravity expressions and the lower-case Dirac matrices 
are those 
appearing in the Matrix Theory expressions. Also, the supersymmetry variation 
parameters in the matrix theory expressions (\ref{raising}) and (\ref{lowering}) 
are related to those in the (\ref{master}) by
\be
\label{etrans}
\e^{11}_- = {1 \over 2^{1 \over 4} } \e^{10}_- \qquad \e^{11}_+ = - { 1 \over 
2^{3 \over 4} } \e^{10}_+
\ee
Finally, to simplify coefficients, it will be convenient to redefine the 
fermionic currents as
\be
\label{strans}
S^I_{new} = 2^{1\over 4} S^I_{old}
\ee
Starting from (\ref{master}), and using (\ref{gtrans}), (\ref{etrans}), and 
(\ref{strans}), it is now straightforward to write down supersymmetry relations 
appropriate for the conventions of this section. As an example, the expressions 
relating the lowest dimension currents $T^{++}$, $S^+_+$, $T^{+i}$, and 
$J^{+ij}$ are
\bea
\delta_+ T^{++} &=& 0 \label{+t++}\\
\delta_- T^{++} &=& k_i \bar{\e}_- \gamma^{0i} S^+_+ \label{-t++}\\
\delta_+ S^+_+ &=& \e_+ T^{++} \label{+s++}\\
\delta_- S^+_+ &=& - \gamma^i \e T^{+i} - {1 \over 2} \gamma^{0ij} \e J^{+ij} 
+ i \slash{k} R_-^- \label{-s++}\\
\delta_+ T^{+i} &=&  {1 \over 4} k_j \bar{\e} \gamma^{0ij} S_+^+ 
\label{+t+i}\\
\delta_+ J^{+ij} &=& - {1 \over 4} k_k \bar{\e} \gamma^{ijk} S_+^+ 
\label{+j+ij}
\eea
Here, we have omitted the auxiliary currents in some of the terms since they 
turn out to be zero. Additional explicit supersymmetry relations involving the 
higher dimensional currents will be given below.

To use these relations, we begin with the expression for $T^{++}$. For the BFSS 
model, this was derived in \cite{kt2} and is given by
\[
T^{++}(k) = \str(e^{ik \cdot X}) \; .
\]
This expression can be motivated by noting that it is the simplest operator 
whose Fourier transform gives a sum of delta-functions for diagonal $X^i$. This 
property is required since $T^{++}$ is exactly the operator measuring the 
spatial 
density of D0-brane charge in the $\a' \to 0$, $g_s \to 0$ limit of type IIA 
string theory. 

Since $T^{++}$ is purely bosonic, the relation (\ref{+t++}) is trivially 
satisfied (and the auxiliary current that could have appeared on the right-hand 
side is determined to be 0.) The first nontrivial relation is (\ref{-t++}). 
Evaluating the left side, we find
\[
\partial_- T^{++} = k_i \bar{\e} \gamma^i \l
\]
This is consistent with the right-hand side if
\[
S^+_+ = \str(\gamma^0 \l e^{i k \cdot X}) \; .
\]
Again, we can take the possible auxiliary current term to vanish. We
can check that this expression satisfies the lowering supersymmetry
relation (\ref{+s++}). Considering now (\ref{-s++}), we find that the
left side gives
\beas
\delta_- S^+_+ &=& \str(e^{i k \cdot X} \{ - \gamma^i \e_- F_{0i} + {1 \over 2} 
\gamma^{0ij} \e_- F_{ij}  + k_i \gamma^0 \l \bl \gamma^i \e_- \} ) \\
&=&  - \gamma^i \e_- \; \str(e^{ik \cdot X} \{ F_{0i} + {1 \over 8} \bl 
\gamma^{0ij} \l k_j \} )\\
&& - {1 \over 2} \gamma^{0ij} \e_- \; \str(e^{i k \cdot X} \{ - F_{ij} + {1 
\over 
8} \bl \gamma^{ijk} \l k_k \} )\\
& & + \slash{k} \; \str(e^{i k \cdot X} \{ - {1 \over 32} \gamma^{jk} \e \bl 
\gamma^{0jk} \l - {1 \over 96}  \gamma^{0jkl} \e \bl \gamma^{jkl} \l \} )
\eeas 
where we have used a Fierz identity 
\[
\l \l^\dagger = {1 \over 96} \gamma^{\mu \nu \l} \gamma^0 \; ( \bl \gamma_{\l 
\nu 
\mu} \l) 
\]
to rearrange the fermion terms
into a form consistent with the right hand side of (\ref{-s++}). It is
clear that the relation (\ref{-s++}) can only be satisfied if we
choose the auxiliary term $R^-_-$ at order $k^0$ to be
\[
R^-_- = {i \over 32} \gamma^{jk} \e \; \str(e^{ik \cdot X}  \bl
\gamma^{0jk} \l )  
+ {i \over 96} \gamma^{0jkl} \e \; \str(e^{ik \cdot X} \bl \gamma^{jkl} \l) 
\]
Assuming no terms in $R^-_-$ at higher order in $k$ (this is
consistent but not obviously necessary), we may conclude that
\be
\label{dim2}
T^{+i} = I_2^{0i} \; , \qquad  J^{+ij} = I_2^{ij}
\ee
where 
\be
\label{cov}
I_2^{\mu \nu} = \str(\{ - F^{\mu \nu} + {1 \over 8} \bl \gamma^{\mu \nu i} \l
k_i \} e^{i k \cdot X})
\ee
These expressions match the results derived originally in 
\cite{kt2,tvm}.\footnote{Actually, the original derivation did not rule out the 
possibility of additional fermion terms beyond order $k$ though this
has been established in \cite{dnp}.}  

As discussed in \cite{tvp}, the fact that $T^{+i}$ and $J^{+ij}$ are different 
components of a single Lorentz covariant expression $I^{\mu \nu}_2$ (where $\mu$ 
and $\nu$ are ten-dimensional indices) is actually required. Using the relations 
(\ref{relationship}) and the T-duality relations (\ref{tduality}), one finds 
that $T^{+i}$ and $J^{+ij}$ 
couple to the $(0i)$ and $(ij)$ components of the NS-NS two-form field in the 
Dp-brane action where $i$ and $j$ are taken along the brane directions. Thus, we 
could have deduced the expression for $T^{+i}$ from $J^{+ij}$ or vice-versa. 
This approach will be quite useful when we derive currents in the 
Berkooz-Douglas model.

Using additional supersymmetry relations, we can proceed in this way
to derive expressions for the higher dimensional currents. Since these
expressions are already known, we will skip the details of the
derivation and present the results. From (\ref{master}),
we find that the relevant supersymmetry relations involving currents up to 
dimension 4 include raising supersymmetry relations  
\bea
\delta_- T^{++} &=& k_i \bar{\e}_- \gamma^{0i} S^+_+ \nonumber \\
\delta_- S^+_+ &=& - \gamma^i \e_- T^{+i} - {1 \over 2} \gamma^{0ij} \e_-
J^{+ij} \nonumber \\
&& + i \slash{k} R_-^- \nonumber \\
\delta_- T^{+i} &=& -{i \over 2} \partial_t S_+^+ \gamma^i \e_- + {1
\over 2} k_j S_+^i \gamma^j \e_- - {\sqrt{2} \over 4} k_j S^+_-
\gamma^{ji} \e_- \nonumber \\
\delta_- J^{+ij} &=& {\sqrt{2} \over 12} \bar{\e}_- \gamma^{ijk} S_-^+
k_k - {1 \over 6} \bar{\e}_- \gamma^{lijk} S_+^l k_k - \bar{\e}_-
\gamma^{[ij} S_+^{k]} k_k \label{bfssrel1}\\
\delta_- S_+^i &=& - \gamma^j \e_- T^{ij} - {1 \over 2} \gamma^{0jk}
\e_- J^{ijk} + \gamma^0 \e_- J^{+-i} - {1 \over 6} \gamma^{jkl} \e_-
M^{+-ijkl}\nonumber \\
&&  - \gamma^i \partial_t R_-^- + \sqrt{2} i \gamma^{ij} k_j R_-^+ -
i \gamma^{ijk} k_j R_k^- \nonumber \\
 \delta_- S^+_- &=& - \sqrt{2} \e_- T^{+-} + \sqrt{2} \gamma^{0i} \e_-
J^{+-i} + {\sqrt{2} \over 4!} \gamma^{ijkl} \e_- M^{+-ijkl} \nonumber \\
& & - i \slash{k} R_-^+ + \sqrt{2} i \gamma^{ij} k_i R_j^-  \nonumber 
\eea
and lowering supersymmetry relations 
\beas
\delta_+ T^{++} &=& 0 \nonumber \\
\delta_+ S^+_+ &=& \e_+ T^{++} \nonumber \\
\delta_+ T^{+i} &=& {1 \over 4} k_j \bar{\e}_+ \gamma^{0ij} S_+^+ \nonumber \\
\delta_+ J^{+ij} &=& - {1 \over 4} k_k \bar{\e}_+ \gamma^{ijk} S_+^+ \nonumber 
\\
\delta_+ S_+^i &=& \e_+ T^{+i} + \gamma^{0j} \e_+ J^{+ij} \nonumber \\
&&  + \sqrt{2} i \gamma^{ij} k_j \tilde{R}_-^+ - i \gamma^{ijk} k_j
\tilde{R}_k^-  \label{bfssrel2} \\
\delta_+ S^+_- &=& { \sqrt{2} \over 2}  \gamma^i \e_+ T^{+i} +
{\sqrt{2} \over 4} \gamma^{0ij} \e_+ J^{+ij} \nonumber \\
& & - i \slash{k} \tilde{R}_-^+ + \sqrt{2} i \gamma^{ij} k_i \tilde{R}_j^-  
\nonumber \\
\delta_+ T^{ij} &=&  {1 \over 2} k_l \bar{\e}_+ \gamma^{0l(i} S_+^{j)} \nonumber 
\\
\delta_+ T^{+-} &=& - { i \over 4} \e_+ \partial_t S_+^+ + {\sqrt{2}
\over 4} k_l \bar{\e}_+ \gamma^{0l} S_-^+ \nonumber 
\eeas
Here, we have included auxiliary current terms only in cases where
they turn out to be non-zero. Using these, we find that the complete expressions 
for the bosonic BFSS currents with dimension 4 and less are given by
\be
\ba{lll}
T^{++} = I_0 & T^{+i} = I_2^{0i} & J^{+ij} = I_2^{ij}\\
T^{ij} = I_4^{ij} & J^{+-i} = I_4^{0i} & T^{+-} = I_4^{00} - {1 \over
4} I_4^\mu {}_\mu\\
J^{ijk} = I_4^{0ijk} & M^{+-ijkl} = I_4^{ijkl} &
\ea
\ee
where 
\begin{eqnarray*}
I_0 &=& \str(e^{i k \cdot X})\\
I_2^{\mu \nu} &=& \str(\{ -F^{\mu \nu} + {1 \over 8} k_i \bl \gamma^{\mu
\nu i} \l \} e^{i k \cdot X} )\\
I_4^{\mu \nu} &=& \str( \{ F^{\mu \l} F_\l {}^\nu - {i \over 2} \bar{\lambda} 
\Gamma^{(\mu} D^{\nu)} \lambda + {1 \over 4} k_i\bl \gamma^{\l (\mu} {}_i \l 
F_\l 
{}^{\nu)} \\
& & \qquad -  { 1 \over 192} k_i k_j \bl \gamma^{i \l (\mu} \l \bl \gamma_\l
{}^{\nu)j} \l \} e^{i k \cdot X}   )\\
I_4^{\mu \nu \l \s} &=& \str(\{ 3 F^{[\mu \nu} F^{\l \s]} 
+ {i \over 4} \bl \gamma^{\mu \nu \l \s \r} D_\r \l - 
{3 \over 4} k_i \bl \gamma^{i [\mu \nu} \l F^{\l \s]} \\
& & \qquad + {3 \over 16} k_i k_j \bl \gamma^{i[\mu \nu} \l \bl 
\gamma^{\l \s] j} \l \} e^{ik \cdot X} )
\eeas
Complete expressions for the fermionic currents with dimension less than four 
are given by
\beas
S_+^+ &=& \str(\gamma^0 \l e^{i k \cdot X})\\
S_+^i &=& \str(\{ \gamma^\mu \l F_{\mu i} - {1 \over 24} \gamma^\mu \l \bl 
\gamma_\mu {}^{ij} \l k_j \} e^{i k \cdot X})\\
S_-^+ &=& \str({\sqrt{2} \over 2} \{ - \gamma^{0i} \l F_{0i} +
 {1 \over 2} \gamma^{ij} \lambda F_{ij} + {1 \over 12} \gamma^{0 \mu} \lambda 
\bl \gamma_\mu {}^{0j} \lambda k_j \} e^{i k \cdot X})\\
\eeas
For each of the supersymmetry relations we have checked, the auxiliary
currents at a given order in $k$ either may be taken to vanish or are
completely fixed by requiring consistency in the structures on each
side of the equation. 

The bosonic terms and two-fermion terms in these expressions were
derived originally in \cite{kt2} and \cite{tvm}, while the higher order
fermion terms were worked out in \cite{dnp,plefka} using supermembrane results.  
In all cases, the expressions above match those previously derived. 
With more work, it should be 
possible using this approach to determine complete expressions for the remaining 
higher dimension currents, for which only the terms with up to two fermions
have been written down previously \cite{kt2,tvm}. (From the form of the 
supersymmetry
relations, it is clear that the dimension $2k$ bosonic currents will
include terms with up to $2k$ fermions, while the dimension $2k + {3 \over 2}$ 
fermionic currents will contain terms with up to $2k + 1$ fermions.)

\section{Dp-D(p+4) systems and the Berkooz-Douglas matrix model}

We have seen that the supersymmetry relations (\ref{master}) provide a powerful 
tool for deriving spacetime currents in Matrix theory. We would now like to 
apply 
these techniques to derive currents in the Berkooz-Douglas matrix
model, and then use these currents to deduce leading terms in the
Dp-D(p+4) brane actions. To prepare 
for this, we review in this section the field content, symmetries, Lagrangian of 
a system of Dp-branes and D(p+4)-branes, and then recall the Berkooz-Douglas 
matrix model which arises as a low-energy, weak-coupling limit of the 0-4 
system. 

\subsection{Dp-D(p+4) field content}

The field content for the various Dp-D(p+4) systems is related by T-duality, 
just as for the case of a single type of D-brane. To describe the field content, 
it is convenient to begin with the D5-D9 system, which has the largest symmetry 
group. In what follows, we will always take $N$ to be the number of Dp-branes 
and $k$ to be the number of D(p+4)-branes. 

\begin{figure}
\label{fields}
\centerline{\epsfysize=2.5truein \epsfbox{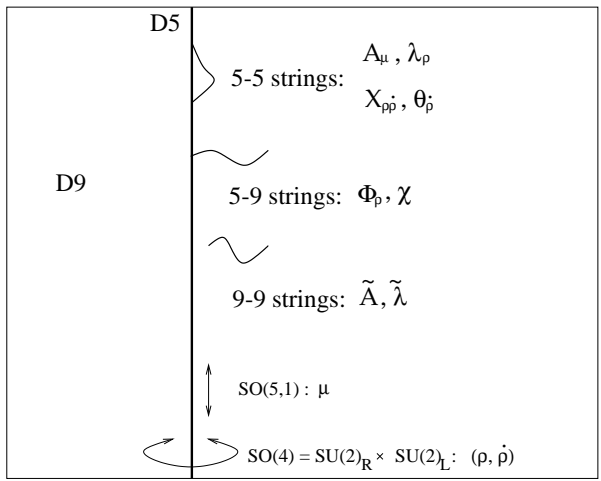}}
 \caption{Summary of field content and symmetries for D5-D9 system.}
\end{figure}

The massless 5-5 and 5-9 strings together give the field content of a 
six-dimensional ${\cal N} = 1$ supersymmetric $U(N)$ Yang-Mills theory, 
consistent 
with the fact that the system of branes preserves 8 supercharges. The 5-5 
strings give rise to the gauge multiplet as well as one adjoint hypermultiplet, 
while the 5-9 strings give $k$ fundamental hypermultiplets. This theory has an 
$SO(4) = SU(2)_L \times SU(2)_R$ internal symmetry related to rotations in the 4 
directions transverse to the 5-brane. To label the fields, we use indices 
$a,b,c,\dots$ to label spacetime indices in the 5-brane directions, lower case 
Greek indices $\a, \r, \s, \dots$ as fundamental $SU(2)_R$ indices and dotted 
lower case greek indices $\da, \dr, \ds, \dots$ as fundamental $SU(2)_L$ 
indices. Then the gauge multiplet fields from the 5-5 strings are a gauge field 
$A_a$ and positive chirality fermion $\lambda_\r$, the adjoint hypermultiplet 
fields from the 5-5 strings are scalars $X_{\r \dr}$ and a negative chirality 
fermion $\t_\dr$, and the fundamental hypermultiplet fields from the 5-9 strings 
are scalars $\P_\r$ and a negative chirality fermion $\chi$. The fermions each 
have 8 real components, because of the Weyl condition as well as 
constraints\footnote{Here $\psi^c \equiv C^{-1} \bar{\psi}^T$ and $C$ is the 
charge conjugation matrix, obeying
\[
C \Gamma^\mu C^{-1} = - (\Gamma^\mu)^T \; , \qquad C^* C = 1 \; , \qquad C^{T} = 
C \; .
\]}
\[
\lambda_\rho = \e_{\r \s} (\lambda^c)^\s \qquad \theta_\dr = -\e_{\dr \ds} 
(\t^c)^\ds
\]
The scalars $X_{\rho \dr}$ also obey a reality condition 
\[
X_{\r \dr} = \e_{\r \s} \e_{\dr \ds} \bar{X}^{\s \ds}
\]
so that they transform as a real vector of $SO(4)$, as desired. 

Finally, the massless 9-9 fields are precisely those of the ${\cal N}=1$ $U(k)$ 
supersymmetric Yang-Mills theory in 10 dimensions, namely a gauge field 
$\tilde{A}_a, \tilde{A}_{\r \dr}$ and a Majorana-Weyl fermion 
$\tilde{\lambda}_\r, \tilde{\lambda}_\dr$. The hypermultiplet from the 5-9 
strings transforms in the antifundamental of the $U(k)$ gauge group. The field 
content is summarized in figure 2. A number of formulae useful for 
manipulating expressions involving six-dimensional spinors and also for relating 
the $SO(4)$ notation to $SU(2) \times SU(2)$ notation are given in appendix B.

\subsection{Dp-D(p+4) action} 

The action and supersymmetry transformation rules for a general six-dimensional 
$N=1$ supersymmetric Yang-Mills theory containing a vector multiplet with 
arbitrary gauge group and an arbitrary set of hypermultiplets is given in 
appendix A. Specializing to the D5-D9 system, we can 
immediately read off the action for the six-dimensional theory describing the 
5-5 and 5-9 strings. The Lagrangian density is given by
\bea
{\cal L} &=& \tr \left(-{1 \over 4} F_{\mu \nu} F^{\mu \nu} - {1 \over 2} D_\mu
\bar{X}^{\rho \dot{\rho}} D^\mu X_{\rho \dot{\rho}} - {i \over 2} \bl^\r 
\gamma^\mu D_\mu \l_\r - {i \over 2} \bt^\da \gamma^\mu D_\mu \t_\da \right) 
\nonumber \\
& & + {\rm tr}\left(- D_\mu \bP^\r D^\mu \P_\r  -i \bc \gamma^\mu D_\mu \chi 
\right) \nonumber \\
& &  + \tr \left( - {1 \over 4} [\bar{X}^{\a \da}, X_{\b \da}][\bar{X}^{\b \db}, 
X_{\a \db}] - \sqrt{2} i \e^{\a \b} \bt^\da [X_{\b \da}, \l_\a] \right) 
\label{pp4flat} \\
&& + {\rm tr} \left( \bP^\a [\bX^{\b \da}, X_{\a \da}] \P_\b +{1 \over 2}\bP^\a 
\P_\b \bP^\b \P_\a - \bP^\a 
\P_\a \bP^\b \P_\b \right. \nonumber \\
& & \left.  + \sqrt{2} i \e^{\a \b} \bc \l_\a \P_\b - \sqrt{2} i \e_{\a \b} 
\bP^\a \bl^\b \c \right) \nonumber 
\eea
Here $\tr$ and ${\rm tr}$ represent traces over $U(N)$ and $U(k)$ indices 
respectively. In addition, we have the usual ten-dimensional ${\cal N}=1$ SYM 
action describing the 9-9 strings, and these couple to the 5-9 strings via the 
covariant derivatives in the action above, which should be defined as 
\[
D_\mu \P_\r = \partial_\mu \P_\r + i A_\mu \P_\r - i \P_\r \tilde{A}_\mu 
\]
where $\tilde{A}_\mu$ represents a pull-back of $\tilde{A}$ to the D5-brane 
worldvolume.

The low-energy actions describing the remaining Dp-D(p+4) systems follow from 
this action by dimensional reduction. The Lagrangian density is then precisely 
the expression above if we define $F_{ab} = i [X^a, X^b]$ and so forth, where 
$a, b, \dots$ are the directions that have been dualized, i.e. the directions 
perpendicular to both sets of branes. 

\subsection{The Berkooz-Douglas matrix model}

The Berkooz Douglas matrix model was proposed as a matrix model for DLCQ 
M-theory with $N$ units of momentum along the longitudinal direction in the 
presence of $k$ longitudinal M5-branes. It is given by the limit of type IIA 
string theory with $N$ D0-branes and $k$ D4-branes with $g_s \sim \e^{3 \over 2} 
\to 0$ and ${\a}' \sim \e \to 0$, keeping only states with finite energy in the 
limit. In particular, the 
only dynamical degrees of freedom which remain in the limit are the lowest 
energy modes of the 0-0 and 0-4 strings. From the 4-4 strings, only zero-modes 
of the scalars $\tilde{X}_a$ survive, and these become parameters describing the 
transverse positions of the M5-branes\footnote{In the paper of Berkooz and 
Douglas, additional fermionic parameters were included to take into account 
different polarization states of the M5-branes generated by the broken 
supersymmetries. However, such polarization states are only distinguishable for 
M5-branes with all of their worldvolume directions compactified, so we believe 
that these fermionic parameters should be omitted here.}   

The Lagrangian for the Berkooz-Douglas model may therefore be read off from 
(\ref{pp4flat})
\beas
\label{Lmm}
{\cal L} &=& \tr \left( {1\over 2} D_0 X^a D_0 X^a + {i \over 2}
\lambda^{\dagger^\rho} D_0 \lambda_\rho + {1 \over 2} D_0 \bar{X}^{\rho
\dot{\rho}} D_0 X_{\rho \dot{\rho}} + {i \over 2} \theta^{\dagger
\dr} D_0 \theta_\dr \right)\\ 
&&+ {\rm tr} \left(D_0 \bP^\r D_0 \P_\r + i \chi^\dagger D_0 \chi \right) + 
{\cal 
L}_{int}
\eeas
where
\beas
{\cal L}_{int} &=& \tr \left({1 \over 4} [X^a, X^b][X^a,X^b] + {1 \over 2} [X^a,
\bar{X}^{\rho \dot{\rho}}][X^a, X_{\rho \dot{\rho}}] - {1 \over 4}
[\bar{X}^{\a \da}, X_{\b \da}][\bar{X}^{\b \db}, X_{\a \db}] \right)\\
&& - {\rm tr} \left( (\bP^{\r} X^a - x^a \bP^{\r})(X^a \P_\r - \P_\r x^a )  
\right)\\
&& + {\rm tr} \left( \bP^\a [\bX^{\b 
\da}, X_{\a \da}] \P_\b + {1 \over 2}\bP^\a \P_\b \bP^\b \P_\a - 
\bP^\a \P_\a \bP^\b \P_\b \right)\\
& & + \tr \left({1 \over 2} \bl^\r \gamma^a [X^a , \l_\r] + {1 \over
2} \bt^\da \gamma^a [X^a, \t_\da] - \sqrt{2} i \e^{\a \b} \bt^\da
[X_{\b \da}, \l_\a] \right) \\
&& + {\rm tr} \left( \bc \gamma^a (X^a \c - \c x^a) + \sqrt{2} i \e^{\a \b} \bc 
\l_\a \P_\b - \sqrt{2} i \e_{\a \b} \bP^\a \bl^\b \c \right)
\eeas
Here, $x^a$ are diagonal matrices whose $k$ elements describe the transverse 
positions of the $k$ M5-branes. Also, $\tr$ denotes a trace over the $U(N)$ 
gauge indices, while ${\rm tr}$ denotes a trace over the global $SU(k)$ indices. 
The terms in the first line are just the usual quartic potential of the BFSS 
model. The term in the second line gives a mass for the fundamental fields 
on the (classical) Coulomb branch of the theory (when the D0-branes are 
separated from the D4-branes). The terms in the third line plus the last term 
in the first line may also be written together as
\[
-{1 \over 2} \tr({\cal D^A} {\cal D^A})
\] 
where
\[
{\cal D^A} = \sigma^A_\r {}^\s ({1 \over 2} [\bX^{\r \dr}, X_{\s \dr}] - \P_\s 
\bP^\r )
\]
Classically, on the Higgs-branch where $X^a=0$, demanding that this potential
vanishes gives the ADHM equations ${\cal D}_A=0$ whose solutions describe 
configurations of $N$ instantons dissolved in the $k$ coincident D4-branes. 

In most of the remainder of this paper, we will set $x^a = 0$ for simplicity, 
corresponding to the case of coincident M5-branes or D(p+4)-branes. However, the 
$x^a$ dependence can generally be restored by the replacements $X^a \P
\to X^a \P - \P x^a$ and $X^a \c \to X^a \c - \c x^a$. 

The Berkooz-Douglas model is invariant under sixteen supersymmetries, since the 
presence of M5-branes breaks half of the supersymmetries of the BFSS theory. Of 
the ``raising'' supersymmetries $S^+_-$, only those with positive eigenvalue for 
$\gamma^{012345}$ are preserved. These are the 8 supersymmetries inherited from 
the six-dimensional ${\cal N}=1$ theory (\ref{pp4flat}), and act as
\bea
\delta_- A_0 &=& -i \bz^\r \gamma_0 \l_\r \nonumber\\
\delta_- X_a &=& -i \bz^\r \gamma^a \l_\r \nonumber \\
\delta_- \l_\r &=& {i \over 2} [X^a, X^b] \gamma^{ab} \z_\r + D_0 X^a
\gamma^{0a} \z_\r - i \z_\a [\bX^{\a \da}, X_{\r \da}] + 2i \z_\a
\P_\r \bP^\a - i \z_\r \P_\a \bP^\a  \nonumber \\ 
\delta_- X_{\r \dr} &=& \sqrt{2} \e_{\r \s} \bz^\s \t_\dr \nonumber \\
\delta_- \t_\dr &=& \sqrt{2} i \e^{\a \b} D_0 X_{\a \dr}  \gamma^0 \z_\b
- \sqrt{2} \e^{\a \b} [X^a, X_{\a \dr}] \gamma^a \zeta_\b \nonumber \\
\delta_- \P_\r &=& \sqrt{2} \e_{\r \s} \bz^\s \c \nonumber\\
\delta_- \c &=& \sqrt{2} i \e^{\a \b} D_0 \P_\a \gamma^0 \z_\b - \sqrt{2}
\e^{\a \b} X^a \P_\a \gamma^a \z_\b \label{BDraise}
\eea
The remaining 8 supersymmetries are the half of the ``lowering'' supersymmetries 
which have a negative eigenvalue for $\gamma^{012345}$. These act only on 
$\theta_\dr$, as
\be
\label{BDlower}
\delta_+ \t_\dr = \gamma_0 \z_\dr
\ee

Choosing the gauge $A_0 = 0$, the Hamiltonian of the theory is given by 
\[
\tr( {1 \over 2} \Pi^a \Pi^a + {1 \over 2} \bar{\Pi}^{\r \dr} \Pi_{\r \dr} ) + 
\bar{P}^\rho P_\rho - {\cal L}_{int}
\]
The quantum commutation relations are given by\footnote{In this formula $\a$ and 
$\b$ indicate spinor indices, which are suppressed in the remainder of the 
paper. Everywhere else, $\a$ and $\b$ will be SU(2) indices. } 
\beas
\left[ X^a_{kl}, \Pi^b_{mn} \right] &=& i \delta_{kn} \delta_{lm} \delta^{ab} \\
\left[ X_{\r \dr \; kl} , \bar{\Pi}^{\s \ds }_{mn} \right] &=& i 
\delta^\s_\r 
\delta^{\ds}_{\dr} \delta_{kn} \delta_{lm} \\
\left[ \P^{iI}_\r, \bar{P}^{\s Jj} \right] &=& i \delta^{IJ}
\delta^{ij} \delta_{\r}^{\s}\\
\left\{ (\l_{\r \; kl})_\a , (\l^{\dagger  \sigma}_{mn})_\b \right\} &=& 
\delta_\r^\s 
\delta_{\a \b} \delta_{kn} \delta_{lm}\\
\left\{ (\theta_{\dr \; kl})_\a , (\t^{\dagger \ds}_{mn})_\b \right\} &=& 
\delta_\dr^\ds \delta_{\a \b} \delta_{kn} \delta_{lm}\\
\left\{ (\c^{iI})_\a , (\c^{\dagger Ij})_\b \right\} &=& \delta^{ij}
\delta^{IJ} \delta_{\a \b} 
\eeas
Variation of the action gives the following equations of motion that will be 
required in deriving the currents
\bea
D_0 D_0 X^a &=& -[[X^a,X^b],X^b] -i \bX^{\r \dr} D_a 
X_{\r \dr} + i D_a X_{\r \dr} \bX^{\r \dr} - i \P_\r D_a \bP^\r +i D_a \P_\r 
\bP^\r \nonumber \\
& & - \bl^\r \gamma^a \l_\r - \bt^\dr \gamma^a \t_\dr - \gamma^a \c \bc  
\nonumber \\
D_0 D_0 X_{\r \dr} &=& D_a D_a X_{\r \dr} + [[\bX^{\b \db}, X_{\r \db}]X_{\b 
\dr}] + 2 [X_{\a \dr} , \P_\r \bP^{\a}] - [X_{\r \dr}, \P_{\a} \bP^{\a}] 
\nonumber \\
& & + \sqrt{2} i \e_{\a \r} \bl^\a \t_\dr - \sqrt{2} i \e_{\da \dr} \bt^{\da} 
\l_\r \label{eqomo} \\
D_0 D_0 \P_\r &=& D_a D_a \P_\r + [\bX^{\b \da}, X_{\r \da}] \P_\b  + \P_\b 
\bP^\b \P_\r -2 \P_\r \bP^\a \P_\a \nonumber \\
& & - \sqrt{2} i \e_{\r \b} \bl^\b \c \nonumber \\
D_0 \l_\r &=& \gamma^{0a} D_a \l_\r - \sqrt{2} \e_{\a \r} \gamma^0 \chi \bP^\a + 
\sqrt{2} \e_{\r \b} [\bX^{\b \dr}, \gamma^0 \t_\dr] + \sqrt{2} \P_\r \gamma^0 
\c^{c} \nonumber \\
D_0 \t_\da &=& \gamma^{0a} D_a \t_\da + \sqrt{2} \e^{\a \b} [X_{\b \da}, 
\gamma^0 
\lambda_\a] \nonumber \\
D_0 \c &=& \gamma^{0a} D_a \c - \sqrt{2} \e^{\a \b} \gamma^0 \l_\a \P_\b 
\nonumber  
\eea
The Gauss Law constraint, arising from the equation of motion for $A_0$ is
\beas
0 &=& D_a F_{0a} +i \bX^{\r \dr} D_0 X_{\r \dr} - i D_0 X_{\r \dr} \bX^{\r \dr} 
+ i \P_\r D_0 \bP^\r - i D_0 \P_\r  \bP^\r\\
& & - \bl^\r \gamma^0 \l_\r - \bt^\dr \gamma^0 \t_\dr - \gamma^0 \c \bc  
\eeas
In the quantum mechanical matrix model, physical states are required
to vanish under the action of this operator (the generator of $U(N)$ 
transformations.)

We now have all the tools needed to derive expressions for the currents in the 
Berkooz-Douglas model.

\section{Currents in the Berkooz-Douglas model}

In this section, we would like to derive expressions for currents in the Berkooz 
Douglas model using the supersymmetry relations (\ref{master}), as we did for 
the BFSS model in section 4. Since the $SO(9)$ symmetry of the BFSS model is now 
broken to $SO(5) \times SO(4) \sim SO(5) \times SU(2)_L \times SU(2)_R$, we 
should further split the currents to reflect transformation properties under 
this smaller group. For example, the membrane current $J^{+ij}$ now splits up 
into $J^{+ab}$, $J^{+a} {}_{\r \dr}$, $J^{+A}$, and $J^{+ \dot{A}}$, where $A$ 
and $\dot{A}$ represent indices in the ${\bf 3}$ of $SU(2)_R$ and $SU(2)_L$ 
respectively (corresponding to the self-dual and anti-self-dual parts of 
$J^{+ij}$ with $i$ and $j$ in transverse M5-brane directions). Our conventions 
for the normalizations all of these components appear in appendix
B. The reader interested only in the results will find these
summarized in section 7.

Unfortunately, unlike the BFSS case, we do not have available to us the complete 
expression for $T^{++}$. A suggestion for the full form of this operator in the 
Berkooz-Douglas theory will be given in section 9, but for the present, we 
will start only with the knowledge that $T^{++}$ (and each of the other 
currents) should reduce to the BFSS 
expression when the fundamental fields are set to zero.\footnote{Physically, 
this requirement is clear since the charge density of a configuration of 
D0-branes very far from the D4-branes should be the same as if the D4-branes 
were not there.} 

Thus, we have
\[
T^{++} = \str(e^{i k \cdot X} \{1 + \dots \})
\]
where now $k \cdot X = k_a X^a + \bar{k}^{\r \dr} X_{\r \dr}$ and the dots 
indicate terms with one or more pairs of fundamental
fields. Fortunately, we will find that these unknown terms are at
least partly determined from the known terms by demanding consistency of the 
supersymmetry relations (\ref{master}). 

To proceed, it is useful to think about the set of unknown terms as an
expansion in the number of powers of momentum. Physically, terms in
the Taylor expansion of the currents in powers of momentum 
correspond to the multipole moments of the current distribution. Thus,
if we can determine the leading terms in this momentum expansion, we
will know the operators measuring the monopole moment, dipole
moment, etc.. of the various currents. The form of the supersymmetry
relations (\ref{master}) imply that the $l$-pole moment of a dimension
$d$ current will be determined in terms of the $(l+1)$-pole and lower
moments of the dimension $(d-2)$ currents. Thus, working with a
partial expression for $T^{++}$ up to order $l$ in momenta, we should
be able to use the supersymmetry relations to determine terms in the
dimension $d$ currents up to order $l-{d \over 2}$ in momenta. 

By dimension counting, the leading unknown terms (bilinear in the
fundamental fields) must have at least two powers of momentum and will
come in at the level of the quadrupole moment of $T^{++}$. Up to order
$k^2$, it is easy to see that $T^{++}$ must take the form
\bea
T^{++} &=& \str(e^{i k \cdot X} \{1 + a k_{\r \dr} \bar{k}^{\r \dr}
\P_\s \bP^\s + b k_a k_a  \P_\s \bP^\s +\dots \}) \label{expand}\\
&=& \str(1 + i k_a X^a + i k_{\r \dr} X^{\r \dr} - {1 \over 2} k_a k_b X^a X^b - 
k_a \bar{k}^{\r \dr} X^a X_{\r \dr} \nonumber\\
&&  \qquad - {1 \over 2} \bar{k}^{\r \dr} \bar{k}^{\s \ds} X_{\r \dr} X_{\s \ds}
  + a k_{\r \dr} \bar{k}^{\r \dr} \P_\s \bP^\s + b k_a k_a  \P_\s 
\bP^\s + {\cal O}(k^3)) \nonumber
\eea
for some coefficients $a$ and $b$. 

This expression will be our starting point in applying the supersymmetry 
relations (\ref{master}). We will first determine the terms in the currents that 
follow from the order $k$ terms in (\ref{expand}), and then determine the 
additional terms following from the order $k^2$ terms in (\ref{expand}). 
Together, these give the monopole moments for all currents of dimension four and 
less, the dipole moments for all currents of dimension two and less, and the 
quadrupole moment of $T^{++}$. 

\subsection{Order $k$}

The order $k$ terms in $T^{++}$ will determine the leading terms in the 
operators of dimension 2 and less. The relevant supersymmetry relations follow 
from those of the BFSS model (\ref{bfssrel1}) and (\ref{bfssrel2}) using the 
relations of appendix B to reduce 
them to the six-dimensional notation. For the raising supersymmetries, we find
\bea
\delta_- T^{++} &=& -k_a \z^{\dagger \r}_- \gamma_a S^+_{+ \r} +
\sqrt{2} i \bar{k}^{\r \dr} \e_{\r \s} \z^{\dagger \sigma}_- S^+_{+
\dr} \nonumber \\
& & + (k_a k_a + \bar{k}^{\r \dr} k_{\r \dr}) R^{++} \label{rt++}\\
\delta_- S^+_{+ \r} &=& - {1 \over 2} \gamma^{0ab} \zeta_\r J^{+ ab} -
\gamma^a \zeta_\r T^{+a} - \sqrt{2} i \gamma^0 \s^A_{\r} {}^\a \z_\a
J^{+ A} \nonumber \\
& & + i \slash{k} R_{- \r}^- + \sqrt{2} \bar{k}^{\s \dr} \e_{\r \s}
R^-_{- \dr}\label{rs+r}\\
\delta_- S^+_{+ \dr} &=& - \sqrt{2} i \e^{\a \b} \z_\b T^+ {}_{\a \dr}
- \sqrt{2} i \gamma^{0a} \e^{\a \b} \z_\b J^{+ a} {}_{\a \dr}
\nonumber \\
& & + i \slash{k} R_{- \dr}^- + \sqrt{2} \e_{\ds \dr} \bar{k}^{\r \ds} 
R^-_{- \r} \label{rs+dr}
\eea
Varying
\[
T^{++} = \str(1 + k_a X^a + k_{\r \dr} X^{\r \dr} + {\cal O}(k))
\]
using the raising supersymmetry transformations (\ref{BDraise}) and applying 
the relation (\ref{rt++}) determines the leading terms in $S^+_{+ \r}$ and 
$S^+_{+ 
\dr}$,
\beas
S^+_{+ \r} &=& \tr (\gamma^0 \l_\r + {\cal O}(k)) \\
S^+_{+ \dr} &=& \tr (\gamma^0 \theta_\dr  + {\cal O}(k))
\eeas
Note that the auxiliary current term in (\ref{rt++}) is only relevant at the 
next order. 

From these fermionic currents, we may now use the relations (\ref{rs+r}) and 
(\ref{rs+dr}) to determine the monopole terms in the dimension 2 currents. 
Again, the auxiliary currents are not relevant at this order, so it is 
straightforward to determine
\beas
T^{+a} &=& \tr(F_{0a} + {\cal O} (k) )\\
T^+ {}_{\r \dr} &=& \tr(D_0 X_{\r \dr} + {\cal O} (k) )\\
J^{+A} &=& \tr({ \sqrt{2} \over 4} F^\a {}_\b \sigma^A_\a {}^\b - {\sqrt{2} 
\over 2} \sigma^{A}_\r {}^\s \P_\s \bP^\r + {\cal O} (k) )\\
J^{+a} {}_{\r \dr} &=& \tr(-D_a X_{\r \dr} + {\cal O} (k) )\\
J^{+ab} &=& \tr(-F^{ab} + {\cal O} (k) ) \; .
\eeas
where we define
\[
F^\a {}_\b \equiv [\bX^{\a \dr}, X_{\b \dr}] \; , 
\]
and $F_{ab}$, $D_a X_{\r \dr}$, etc... are to be understood as dimensional 
reduced six-dimensional expressions.

There is one additional dimension 2 current, $J^{+ \dot{A}}$ which does not 
appear in the supersymmetry relations (\ref{rs+r}) or (\ref{rs+dr}). 
At leading order, there are no possible terms involving the fundamental fields 
that are consistent with the transformation properties, so the monopole term 
may 
be read off from the appropriate components of the BFSS current $J^{+ij}$. In 
the six-dimensional language, we find
\[
J^{+ \dot{A}} = \tr( { \sqrt{2} \over 4} F^\da {}_\db \sigma^{\dot{A}}_\da 
{}^\db + {\cal O} (k) )
\]
where 
\[
F^\da {}_\db \equiv [\bX^{\r \da}, X_{\r \db}] \; , 
\]
So far, the only new term involving the fundamental fields appears in the 
current $J^{+A}$ (transforming in the {\bf 3} of $SU(2)_R$). This operator 
(which couples to a constant 3-form potential $C_{+A}$) represents a FI-term 
deformation of the matrix model preserving all 16 supersymmetries. The matrix 
model with this deformation corresponds to ``light-like'' noncommutativity on 
the M5-branes and was discussed originally in \cite{abs}.

\subsection{Order $k^2$}

Having determined all terms that follow from the order $k$ terms in $T^{++}$ 
we now turn to order $k^2$, which becomes substantially more involved.

Our starting point is
\[
T^{++} = \str(e^{i k \cdot X} \{1 + a k_{\r \dr} \bar{k}^{\r \dr}
\P_\s \bP^\s + b k_a k_a  \P_\s \bP^\s + {\cal O}(k^3) \} )
\]
Varying under the supersymmetry transformations (\ref{BDraise}) and applying the 
relation (\ref{rt++}), we find at this order the possibility of an auxiliary 
term
\[
R^{++} = \alpha \z^{\dagger \r}_- \tr ( \gamma^0 \chi \bP^\s \e_{\r \s}
+ \P_\r \gamma^0 \c^c )
\]
with an unknown coefficient $\alpha$. Taking into account this term, the 
dimension ${3 \over 2}$ fermionic currents to order $k$ are then determined to 
be
\beas
S^+_{+ \r} &=& \str \left( e^{ik \cdot X} \left\{ \gamma^0 \l_\r + ( b
\sqrt{2} + \a) k_a (\e_{\s \r} \gamma^{0a} \c \bP^{\s} - \P_\r
\gamma^{0a} \c^c) + {\cal O}(k^2) \right\} \right)\\
S^+_{+ \dr} &=& \str \left( e^{ik \cdot X} \left\{ 
\gamma^0 \theta_\dr - \sqrt{2} i (\sqrt{2} a + \alpha) k_{\r \dr}
(\gamma^0 \chi \bP^\rho - \e^{\r \s} \P_\s \gamma^0 \chi^c ) + {\cal O}(k^2)
\right\} \right)
\eeas
Next, we vary these currents under the raising supersymmetries (\ref{BDraise}) 
and use the relations (\ref{rs+r}) and (\ref{rs+dr}). At this order, there are 
many possible terms that might appear in the auxiliary currents $R^-_- {}_\r$ 
and $R^-_- {}_\dr$, however, it turns out that they are all determined uniquely. 
Firstly, there are terms involving only the adjoint fields that appeared 
previously in our BFSS calculation as $R^-_-$,
\beas
(R_-^- {}_\r)_1 &=& \tr( -{i \over 4} \z_\b \bl^\b \gamma^0 \l_\r - {i \over 4} 
\gamma^{0a} \z_\b \bl^\b \gamma^a \l_\r + {i \over 16} \gamma^{ab} \z_\r \bt^\dr 
\gamma^{0ab} \t_\dr )\\
(R^-_- {}_\dr)_1 &=& \tr({i \over 4} \gamma^0 \zeta_\r \bl^\r \t_\dr - {i \over 
4} \gamma^a \zeta_\r \bl^\r \gamma^{0a} \t_\dr - {i \over 8} \gamma^{0ab} \z_\r 
\bl^\r \gamma^{ab} \t_\dr )
\eeas
In addition, the following terms involving the fundamentals are allowed by 
symmetry
\bea
(R^-_- {}_\r)_2 &=& {\rm tr} ( \a_1 \z_\r \bP^\s \dot{\P}_\s + \a_2 \z_\b \bP^\b 
\dot{\P}_\r + 
\a_3 \z_\r \dot{\bP}^\s \P_\s + \a_4 \z_\b \dot{\bP}^\b \P_\r \nonumber \\
& & + \b_1 \gamma^{0a} \z_\r \bP^\s X^a \P_\s + \b_2 \gamma^{0a} \z_\b \bP^\b 
X^a \P_\r  \label{auxil1} \\      
&& + \eta_1 \z_\r \bc \gamma^0 \c + \eta_2 \gamma^{0a} \z_\r \bc \gamma^a \c + 
\eta_3 \gamma^{ab} \zeta_\r \bc \gamma^{0ab} \c ) \nonumber \\ 
(R^-_- {}_\dr)_2 &=& {\rm tr} (\mu_1 \e^{\a \b} \gamma^0 \z_\r \bP^\r X_{\b \dr} 
\P_\a + \mu_2 \e^{\r \b} \gamma^0 \z_\r \bP^\s  X_{\b \dr}\P_\s ) \nonumber 
\eea
The undetermined coefficients are constrained by the requirement that 
(\ref{rs+r}) and (\ref{rs+dr}) are consistent, that is, the supersymmetry 
variation on the left hand side minus the auxiliary current terms on the right 
hand side must give a set of terms of the same form as the unknown terms on the 
right hand side. Additional constraints are provided by conservation laws for 
the stress energy tensor and membrane current,\footnote{See appendix C
for a discussion of the conserved currents in the Berkooz-Douglas model.} 
\beas
\dot{T}^{++} &=& i k_a T^{+a} + i\bar{k}^{\r \dr} T^{+} {}_{\r \dr}\\
\dot{J}^{+a} {}_{\r \dr} &=& i k_b J^{ba} {}_{\r \dr} - {\sqrt{2}
\over 2} k_{\s \dr} \s^A_\r {}^\s J^{aA} - {\sqrt{2}
\over 2} k_{\r \ds} \s^{\dot{A}}_\dr {}^\ds J^{a \dot{A}} \; .
\eeas
The first equation places obvious constraints on $T^{+a}$ and $T^{+} {}_{\r 
\dr}$, while the antisymmetry in $a$ and $b$ of $J^{ba} {}_{\r \dr}$ on the 
right hand side of the second equation forbids terms in $J^{+a} {}_{\r \dr}$ 
at order $k$ proportional to $k^a$. 

Using the constraints provided by the consistency of the supersymmetry relations 
and the two conservation laws, the undetermined coefficients in the auxiliary 
currents are determined to be
\bea
\a_1 &=& (b + \sqrt{2} \a), \qquad \a_2 = -2(b + \a {\sqrt{2} \over 2}) ,  
\qquad \a_3 = -b, \qquad \a_4 = 2(b+ {\sqrt{2} \over 2} \a) \nonumber \\ 
\b_1 &=& -2i(b+ {\sqrt{2} \over 2} \a), \qquad \b_2 = 4i(b + {\sqrt{2} \over 2} 
\a) \label{auxil2}\\
\eta_3 &=& -{i \over 2} (a + {\sqrt{2} \over 2} \a), \qquad 
\eta_1 = \eta_2 = \mu_i = 0 \nonumber 
\eea
The dimension 2 currents are now determined up to order $k$ from (\ref{rs+r}) 
and (\ref{rs+dr}) to be
\bea
T^{+a} &=& \str ( e^{ik \cdot X} \{ 
D_0 X^a + {1 \over 8} \bl^\b \gamma^{0ab} \l_\b k_b + {1 \over 8}
\bt^\dr \gamma^{0ab} \t_\dr k_b + {\sqrt{2} \over 4} i \bar{k}^{\s
\ds} \e_{\s \r} \bl^\r \gamma^{0a} \t_\ds \nonumber \\
&& \qquad \qquad -ib k_a (\P_\s \dot{\bP}^\s +
\dot{\P}_\s \bP^\s ) + (b-a) k_b \bc^c \gamma^{0ab} \c^c \} ) \nonumber \\
T^+ {}_{\r \dr} &=& \str ( e^{ik \cdot X} \{
D_0 X_{\r \dr} - { \sqrt{2} \over 4} i \e_{\r \b} \bl^\b
\gamma^{0 a} \t_\dr k_a - {1 \over 4} \bt^\ds \gamma^{0} \t_\dr 
k_{\r \ds} - {1 \over 4} \bl^\s \gamma^0 \l_\r k_{\s \dr} \nonumber \\
&& - bi k_{\r \dr} \dot{\P}_\s \bP^\s + i (b-2a) 
k_{\r \dr} \P_\s \dot{\bP}^\s +2i(b-a) k_{\s \dr} \dot{\P}_\r \bP^\s + 2i (a-b) 
k_{\s \dr} \P_\r \dot{\bP}^\s \} ) \nonumber \\
J^{+A} &=& \str( e^{ik \cdot X} \{
{ \sqrt{2} \over 4} F^\a {}_\b \sigma^A_\a {}^\b - {\sqrt{2} \over 2}
\sigma^{A}_\r {}^\s \P_\s \bP^\r + {i \sqrt{2} \over
8} \bl^\r \sigma^A_\r {}^\s \gamma^a \l_\s k_a + {1 \over 4} \bar{k}^{\r \dr}
\sigma^A_\r {}^\s \e_{\s \tau} \bl^\tau \t_\dr \} ) \nonumber \\
J^{+a} {}_{\r \dr} &=& \str ( e^{ik \cdot X} \{ 
-i[X_a, X_{\r \dr}] - {\sqrt{2} \over 4} i \e_{\r \b} \bl^{\b}
\gamma^{ab} \t_\dr k_b - {1 \over 4} \bt^\ds \gamma^a \t_\dr k_{\r \ds}
- {1 \over 4} \bl^\s \gamma^a \l_\r k_{\s \dr} \nonumber \\
&& \qquad \qquad + 4(b-a) k_{\b \dr} X_a \P_\r \bP^\b + 2
(a-b) k_{\r \dr} \P_\s \bP^\s X^a 
  \} ) \nonumber \\
J^{+ab} &=& \str ( e^{ik \cdot X} \{ 
-F^{ab} + {1 \over 8} \bl^\r \gamma^{abc} \l_\r k_c + {1 \over 8}
\bt^\dr \gamma^{abc} \t_\dr k_c \label{dim2currents} \\ && \qquad \qquad 
\qquad + {\sqrt{2} \over 4} i \bar{k}^{\s \ds} \e_{\s \r} \bl^\r \gamma^{ab} 
\t_\ds + 
(b-a) k_c \bc \gamma^{abc} \c \} ) \nonumber 
\eea
It is interesting to note that all dependence on the coefficient $\alpha$ of 
$R^{++}$ has cancelled in these expressions. 

As we noted earlier, the remaining dimension 2 current $J^{+ \dot{A}}$ does not 
appear in the supersymmetry relations (\ref{rs+r}) and (\ref{rs+dr}). The terms 
involving only the adjoint fields are determined from the BFSS current 
$J^{+ij}$, but there are additional possible terms involving the fundamentals. 
The most general possibility consistent with the symmetries is 
\beas
J^{+ \dot{A}} &=& \str ( e^{ik \cdot X} \{ 
{\sqrt{2} \over 4} F^\da {}_\db \sigma^{\dot{A}}_\da {}^\db 
+ {i \sqrt{2} \over 8} \bt^\dr \sigma^{\dot{A}}_\dr {}^\ds \gamma^a \t_\ds k_a - 
{1 \over 4} \e_{\ds \da} \bar{k}^{\a \dr}
\sigma^{\dot{A}}_\dr {}^\ds \bt^\da \l_\a  \\
&& \hspace{1in}  + (A k_{\s \ds} \sigma^{\dot{A}}_\dr
{}^\ds \bX^{\r \dr}
\P_\r \bP^\s + B k_{\s \ds}  \sigma^{\dot{A}}_\dr {}^\ds \bX^{\s \dr} \P_\r 
\bP^\r) \} ) \; .
\eeas
However, it must be that $A=B=0$ since the position space current must take the 
form $J(X_{\r \dr} - x_{\r \dr})$ by translation invariance. In momentum space, 
this implies $J(k,X_{\r \dr} + x_{\r \dr}) = e^{i k \cdot x} J(k,X)$, and the 
extra terms above do not satisfy this.

The next set of currents are the fermionic currents of dimension 5/2. Since 
$(S^+_+)_\dr$ is a conserved current (it generates the lowering supersymmetry) 
it obeys the conservation relation
\[
\dot{S}^+_{+ \dr} = i k_a (S_+^a)_\dr + i \bar{k}^{\s \ds} (S_{+ \s
\ds})_\dr
\]
Inserting the expression for $(S^+_+)_\dr$ above on the left side, we find
\beas
(S_+^a)_\dr &=& \str(e^{i k \cdot X} \{ \gamma^0 \t_\dr D_0 X^a 
+ \gamma^b \t_\dr F_{ba} + \sqrt{2} i \e^{\a \b} \l_\a D_a X_{\b \dr} 
+ {\cal O} (k) \}) \\
(S_{+ \s \ds})_\dr &=& \str(e^{i k \cdot X} \{ \gamma^0 \t_\dr D_0
X_{\s \ds} + \gamma^a \t_\dr D_a X_{\s \ds} + \sqrt{2} \e^{\a \b}
\l_\a [X_{\b \dr} , X_{\s \ds}] \\
& & - \sqrt{2} (\sqrt{2} a + \a) ( \e_{\r \s} \e_{\dr \ds} \gamma^0 \chi
D_0 \bP^\r - \e_{\dr \ds} D_0 \P_\s \gamma^0 \c^c - \e_{\r \s}
\e_{\dr \ds} \gamma^a D_a \c \bP^\r \\ 
&& + \e_{\dr \ds}  \P_\s \gamma^a D_a \chi^c
 + \sqrt{2} \e_{\dr \ds} \P_\s \bP^\r \lambda_\r 
+ \sqrt{2} \e_{\dr \ds} \l_\r \P_\s \bP^\r - \sqrt{2} \e_{\dr \ds}
\l_\s \P_\r \bP^\r ) + {\cal O} (k) \} )
\eeas
To arrive at these results, it is necessary to use the equations of motion 
(\ref{eqomo}). 

The leading ($k=0$) term in the current $(S_-^+)_\r$ is the generator of the 
raising supersymmetry. In order to reproduce the transformation rules 
(\ref{BDraise}), we must have
\beas
(S_-^+)_\r &=& \str(e^{i k \cdot X} \{ - {\sqrt{2} \over 2}
\gamma^{0a} \l_\r D_0 X^a - i \e_{\a \r} \gamma^0 \t_\dr D_0 \bX^{\a
\dr} + i \e_{\r \a} \gamma^a \t_\dr D_a \bX^{\a \dr} + {\sqrt{2} \over 4} 
\gamma^{ab} \l_\r F_{ab}\\
 & & - { i \sqrt{2} \over 2} \l_\a F^\a {}_\r -i \e_{\a \r} \gamma^0 \c D_0 
\bP^\a + i D_0 \P_\r \gamma^0 \c^c - \e_{\r \a} \gamma^a \c \bP^\a X_a\\ 
& & + X_a \P_\r \gamma^a \c^c + \sqrt{2} i \P_\r \bP^\a \l_\a - {i
\sqrt{2} \over 2} \P_\a \bP^\a \l_\r + {\cal O}(k) \})
\eeas
where the normalization may be fixed by comparing the terms involving
only adjoint fields with the BFSS current $S_-^+$ using the
formulae of appendix B.

To determine the currents $(S_{+ \s \ds})_\r$ and $(S_-^+)_\dr$, we
can again determine the adjoint field terms directly from the BFSS
currents, while the only fundamental field terms allowed by symmetries
may be fixed easily using the lowering supersymmetry relations
\beas
\delta_+ (S^+_-)_\dr &=& {\sqrt{2} \over 2} \gamma^a \z_\dr T^{+a} + i
\s^{\dot{A}}_\dr {}^\ds \gamma^0 \z_\ds J^{+ \dot{A}} + {\sqrt{2}
\over 4} \gamma^{0ab} \zeta_\dr J^{+ab}\\  
\delta_+ (S_{+ \s \ds})_\r &=& \e_{\r\b} \s^A_\s {}^\b \z_\ds J^{+A} +
\e_{\r \s} \s^{\dot{A}}_\ds {}^\dr \z_\dr J^{+ \dot{A}} \; .
\eeas
We find
\beas
(S_{+ \s \ds})_\r &=& \str(e^{i k \cdot X} \{ \gamma^\mu \lambda_\r 
D_\mu X_{\s \ds} + {\sqrt{2} \over 2} \e_{\r \tau} \t_\ds F^\tau {}_\s
+ {\sqrt{2} \over 2} \e_{\r \s} \t_\dr F^\dr {}_\ds\\ 
& & \qquad \qquad - {\sqrt{2} \over 2}  \e_{\s \tau} \t_\ds \P_\r \bP^\tau 
- {\sqrt{2} \over 2} \e_{\r \tau} \t_\ds \P_\s \bP^\tau + {\cal O}(k) \} ) \\
(S_-^+)_\dr &=& \str(e^{i k \cdot X} \{ - {\sqrt{2} \over 2}
\gamma^{0a} \t_\dr D_0 X^a + i \e_{\da \dr} \gamma^0 \l_\r D_0 \bX^{\r
\da} - i \e_{\ds \dr} \gamma^a \l_\s D_a \bX^{\s \ds}\\
& & \qquad \qquad + {\sqrt{2} \over 4} \gamma^{ab} \t_\dr F_{ab} - { i \sqrt{2} 
\over
2} \t_\ds F^\ds {}_\dr + {\cal O}(k) \})  
\eeas
The remaining current at dimension ${5 \over 2}$ is given by 
\beas
(S_+^a)_\r &=& \str(e^{i k \cdot X} \{\gamma^\mu \l_\r F_{\mu a} 
- \sqrt{2} i \t_\dr \e^{\dr \ds} D_a X_{\r \ds} + \sqrt{2} i \e_{\r \s} 
\chi D_a \bP^\s + \sqrt{2} i  D_a \P_\r \chi^c  \\ 
& & - i (\sqrt{2} a + \a) ( \e_{\s \r} \gamma^0 \chi
D_0 \bP^\s -  D_0 \P_\r \gamma^0 \c^c - \e_{\s \r}
 \gamma^a D_a \c \bP^\s \\ && +  \P_\r \gamma^a D_a \chi^c
 + \sqrt{2}  \P_\r \bP^\s \lambda_\s + \sqrt{2} \l_\s \P_\r \bP^\s - \sqrt{2} 
\l_\r \P_\s \bP^\s ) + {\cal O} (k) \} )
\eeas
This may be determined from the dimension 2 bosonic currents using a raising 
supersymmetry relation, but in practice, we have determined it from the 
dimension 4 currents and the relation (\ref{raisex}) below (we will see that the 
dimension 4 currents will be determined without using this expression and we 
have provided it for completeness). 

We now turn to the dimension 4 bosonic currents. Here, we expect the leading 
($k=0$) terms to be determined from the lower dimensional currents above using 
the supersymmetry relations. For simplicity, we will focus on the purely bosonic 
terms in these currents. The relevant relations are  
\bea
\delta_- (S_+^a)_\r &=& - \gamma^b \z_\r T^{ab} - {1 \over 2} \gamma^{0bc} \z_\r 
J^{abc} - \sqrt{2} i \s^A_\r {}^\s \gamma^0 \z_\s J^{aA} \label{raisex} \\
&& + \gamma^0 \e_\r J^{+-a} - {1 \over 6} \gamma^{bcd} \e_\r M^{+-abcd} - 
\sqrt{2} i \s^A_\r {}^\s \gamma^a \z_\s M^{+-abA} - \gamma^a \partial_t (R_-^-
)_\r \nonumber \\
\delta_- (S^a_+)_\dr &=& -\sqrt{2} i \e^{\a \b} \z_\b (T^a {}_{\a \dr}
+ M^{+-a} {}_{\a \dr}) - \sqrt{2} i \e^{\a \b} \gamma^{0 b} \z_\b 
J^{ab} {}_{\a \dr} \nonumber \\
& & -
{\sqrt{2} \over 2}i  \e^{\a \b} \gamma^{bc} \z_\b M^{+-abc} {}_{\a
\dr} - \gamma^a \partial_t (R_-^-)_\dr \nonumber \\  
\delta_- (S_{+ \r \dr})_\s &=& - \gamma^b \z_\s T^b {}_{\r \dr} - {1 \over 2} 
\gamma^{0bc} \z_\s J^{bc} {}_{\r \dr} + \gamma^0 \z_\s J^{+-} {}_{\r \dr} + 
{1 \over 6} \gamma^{bcd} \z_\s M^{+-bcd} {}_{\r \dr} \nonumber \\
& & + 2 \gamma^0 \z_\r J_{\s \dr} - \gamma^0 \z_\s J_{\r \dr} -2 \gamma^b \z_\r 
M^{+-b} {}_{\s \dr} + \gamma^b \z_\s M^{+-b} {}_{\r \dr} + \sqrt{2} i
\e_{\s \r} \partial_t (R^-_-)_\dr \nonumber \\  
\delta_- (S_{+ \r \dr})_\ds &=& \e_{\dr \ds} \s^A_\r {}^\b \gamma^{0a}
\z_\b J^{a A} + \e_{\da \ds} \s^{\dot{A}}_\dr {}^\da \gamma^{0a} \z_\r J^{a 
\dot{A}} -
\sqrt{2} i \e^{\a \b} \z_\b T_{\r \dr \; \a \ds} - \sqrt{2} i \e_{\dr
\ds} \z_\r M^{+-} \nonumber \\
& & - {1 \over 2} \e_{\dr \ds} \s^A_\r {}^\b \gamma^{bc} \z_\b
M^{+-bcA} - { 1 
\over 2} \e_{\db \ds} \s^{\dot{A}}_\dr {}^\db \gamma^{bc} \z_\r M^{+-bc \dot{A}} 
+ \sqrt{2} i \e_{\dr \ds} \partial_t (R^-_-)_\r \nonumber \\
\delta_- (S_-^+)_\r &=& - \sqrt{2} \z_\r (T^{+-} - M^{+-}) + \sqrt{2} 
\gamma^{0a} 
\z_\r J^{+-a} + {\sqrt{2} \over 4!} \gamma^{abcd} \z_\r M^{+-abcd} \nonumber \\
& &   + i \s^A_\r {}^\s \gamma^{ab} \z_\s M^{+-abA} \nonumber \\
\delta_- ({S_-^+})_\dr &=& 2i \e^{\a \b} \gamma^0 \zeta_\b J^{+-} {}_{\a \dr} 
- { i \over 3} \e^{\a \b} \gamma^{abc} \z_\b M^{+-abc} {}_{\a \dr} + 2i \e^{\r 
\a} \gamma^a \zeta_\a M^{+-a} {}_{\r \dr}   \nonumber 
\eea
Here, we have included only auxiliary terms that give a contribution at order 
$k^0$. All of these involve expressions that have already been 
determined, so these relations completely determine the dimension 4 currents. In 
practice, we do not need to use all of these relations, since many of the 
currents may be determined more simply using symmetries and conservation laws.

Using the $\delta_- (S_{+ \r \dr})_\s$ relation, we find 
\beas 
T^a {}_{\r \dr} &=& \str(e^{i k \cdot X} \{ F^{a \mu} D_\mu X_{\r \dr}
+ {i \over 2} D_a X_{\b \dr} F^\b {}_\r + {i \over 2} D_a X_{\r \db}
F^\db {}_\dr\\ 
& & \qquad \qquad - i D_a X_{\a \dr} \P_\r \bP^\a +
 {i \over 2} D_a X_{\r \dr} \P_\a \bP^\a + {\cal O}(k) \} )\\
J^{ab} {}_{\r \dr} &=& \str(e^{i k \cdot X} \{ -D_0 X_{\r \dr} F_{ab}
- F_{0a} D_b X_{\r \dr} + F_{0b} D_a X_{\r \dr} + {\cal O}(k) \} )\\
J^{+-} {}_{\r \dr} &=& \str(e^{i k \cdot X} \{F^{0 \mu} D_\mu X_{\r
\dr} - {i \over 2} F^\s {}_\r D_0 X_{\s \dr} - { i \over 2} F^\ds
{}_\dr D_0 X_{\r \ds} \\
& & \qquad \qquad + i D_0 X_{\a \dr} \P_\r \bP^\a - {i \over 2} D_0 X_{\r
\dr}  \P_\a \bP^\a + {\cal O}(k) \} )\\
M^{+-abc} {}_{\r \dr}  &=& \str(e^{i k \cdot X} \{3 F_{[ab} D_{c]} X_{\r
\dr}  + {\cal O}(k) \} )\\
J_{\r \dr}  &=& \str(e^{i k \cdot X} \{  - {i \over 6} (D_0 X_{\s \dr} 
F^\s {}_\r - D_0 X_{\r \ds} F^\ds {}_\dr)\\
&& \qquad \qquad  + i D_0 X_{\a \dr} \P_\r \bP^\a - {i \over 2} D_0 X_{\r \dr} 
\P_\a \bP^\a + {\cal O}(k) \} )\\
M^{+-a} {}_{\r \dr}  &=& \str(e^{i k \cdot X} \{  {i \over 6}
(D_a X_{\s \dr} F^\s {}_\r - D_a X_{\r \ds} F^\ds {}_\dr)\\
&& \qquad \qquad -i  D_a X_{\a \dr} \P_\r \bP^\a + {i \over 2}  D_a X_{\r \dr} 
\P_\a \bP^\a + {\cal O}(k) \} )
\eeas
The current $T^{+-}$ is the conserved current that generates translations in the 
light-cone time direction, so $T^{+-}(k=0)$ is simply the Matrix model  
Hamiltonian, 
\beas
T^{+-} &=& \str(e^{i k \cdot X} \{ -F_{0 \mu} F^{0 \mu} + D_0 \bX^{\r
\dr} D_0 X_{\r \dr} + 2 D_0 \P_\r D_0 \bP^\r \\
& & + {1 \over 4} F_{\mu \nu} F^{\mu \nu} + {1 \over 2} D_\mu \bX^{\r
\dr} D^{\mu} X_{\r \dr} + D^\mu \P_\r D_\mu \bP^{\r} \\
& & + {1 \over 4} F^\a {}_\b F^\b {}_\a - F^\b {}_\a \P_\b \bP^\a -
{1 \over 2} \P_\b \bP^\b \P_\a \bP^\a +  \P_\a \bP^\b \P_\b \bP^{\a}+
 {\cal O}(k) \} )
\eeas
From appendix C, the dimension 2 currents $T^+ {}_{\r \dr}$ and $J^{+a} {}_{\r 
\dr}$ are conserved and thus obey conservation relations
\beas
\dot{T}^+ {}_{\r \dr} &=& i k_a T^a {}_{\r \dr} + i \bar{k}^{\s \ds} T_{\s 
\ds \; \r \dr}\\
\dot{J}^{+a} {}_{\r \dr} &=& i k_b J^{ba} {}_{\r \dr} - {\sqrt{2} \over 2}  
k_{\s \dr} \s^A_\r {}^\s J^{a A} - {\sqrt{2} \over 2}  k_{\r \ds} 
\s^{\dot{A}}_\dr {}^\ds J^{a \dot{A}}
\eeas
Inserting our expression for $T^+ {}_{\r \dr}$ on the left hand side of the 
first expression, one finds that the resulting expression is only consistent 
with the right hand side (in particular, with a symmetric stress tensor $T_{\s 
\ds \; \r \dr}$ if
\[
a - b = - {1 \over 4}  
\]
where $a$ and $b$ were the undetermined coefficients in $T^{++}$. With this 
restriction, the currents on the right side are determined to be
\beas
T_{\r \dr \; \s \ds} &=& \str(e^{i k \cdot X} \{2a \e_{\r \s} \e_{\dr
\ds} D^\mu \P_\a D_\mu \bP^\a - D_\mu X_{\r \dr} D^\mu X_{\s \ds}  
- {1 \over 8} \e_{\r \s}
\e_{\dr \ds} (F^\a {}_\b F^\b {}_\a + F^\da {}_\db F^\db {}_\da ) \\
&& - { 1 \over 2} \e_{\s \a} \e_{\ds \da} F^\a {}_\r F^\da {}_\dr 
 + (1 - 2b) \e_{\r \s} \e_{\dr \ds} F^\b {}_\a \P_\b \bP^\a 
+ {1 \over 2} \e_{\r \a} \e_{\dr \db} F^\db {}_\ds \P_\s \bP^\a 
\\ && + {1 \over 2} \e_{\s \a} \e_{\ds \db} F^\db {}_\dr \P_\r \bP^\a + 4a 
\e_{\r \s} \e_{\dr \ds} (\P_\a \bP^\s \P_\s \bP^\a 
- {1 \over 2} \P_\b \bP^\b \P_\s \bP^\s) + {\cal O}(k) \} )\\
J^{ab} {}_{\r \dr} &=& \str(e^{i k \cdot X} \{ -D_0 X_{\r \dr} F_{ab}
- F_{0a} D_b X_{\r \dr} + F_{0b} D_a X_{\r \dr} + {\cal O}(k) \} )\\
J^{aA} &=& \str(e^{i k \cdot X} {i \sqrt{2} \over 4} \sigma^A_\s {}^\r 
\{  2 D_0 \bX^{\s \ds} D_a X_{\r \ds} - i F_{0a} F^\s {}_\r \\
&& + 8(b-a) (D_a \P_\r D_0 \bP^\s - D_0 \P_\r D_a \bP^\s + i 
F_{0a} \P_\r \bP^\s) + {\cal O}(k) \} )\\
J^{a \dot{A}} &=& \str(e^{i k \cdot X} {i \sqrt{2} \over 4} 
\sigma^{\dot{A}}_\ds {}^\dr \{ 2 D_0 \bX^{\s \ds} D_a X_{\s \dr} 
- i F_{0a} F^\ds {}_\dr  + {\cal O}(k) \} )
\eeas
Another set of currents may be determined directly from the BFSS currents since 
there are no dimension 4 terms involving fundamental fields consistent with 
their transformation properties. These are
\beas
J^{abc} &=& \str(e^{i k \cdot X} \{3 F^{0[a} F^{bc]} + {\cal O}(k) \} )\\
M^{+-abcd} &=& \str(e^{i k \cdot X} \{3 F^{[ab} F^{cd]} + {\cal O}(k) \} )\\
M^{+-ab \dot{A}} &=& \str(e^{i k \cdot X} {i \sqrt{2} \over 4}
\sigma^{\dot{A}}_\ds {}^\dr \{  -2 D_a \bX^{\s \ds} D_b X_{\s \dr} 
+ i F_{ab} F^\ds {}_\dr  + {\cal O}(k) \} )\\
M^{+-} &=& \str( e^{ i k \cdot X} \{ {1 \over 8} F^\s {}_\r F^\r {}_\s 
- {1 \over 8} F^\dr {}_\ds F^\ds {}_\dr + {\cal O} (k) \} ) 
\eeas
The remaining dimension 4 currents are determined from those we 
already have by relations which follow from $SO(5,1)$ Lorentz 
invariance of the dual D5-D9 system, similar to the relation (\ref{dim2}) 
discussed above 
for the BFSS currents. These imply that $M^{+-abA} = I^{abA}$ while $J^{aA} = 
I^{0aA}$ where $I^{\mu \nu A}$ is the dimensional reduction of a 5+1 dimensional 
Lorentz invariant current. From the expression we have derived for $J^{aA}$ we 
may deduce that 
\beas
I^{\mu \nu A} = I_4^{\mu \nu A} &=& \str(e^{i k \cdot X} {i \sqrt{2} \over 4} 
\sigma^A_\s {}^\r\{ -2 D^\mu \bX^{\s \ds} D^\nu X_{\r \ds} + i F^{\mu \nu} 
F^\s {}_\r \\
&& - 8 (b-a) (D^\nu \P_\r D^\mu \bP^\s - D^\mu \P_\r D^\nu \bP^\s  + i \P_\r 
\bP^\s F^{\mu \nu}) + {\cal O}(k) \} )
\eeas
so that
\beas
M^{+-abA} &=& \str(e^{i k \cdot X} {i \sqrt{2} \over 4} \sigma^A_\s
{}^\r\{  -2 D_a \bX^{\s \ds} D_b X_{\r \ds} + i F_{ab} F^\s {}_\r \\
&& - 8(b-a) (D_b \P_\r D_a \bP^\s - D_a \P_\r D_b \bP^\s + i 
F_{ab} \P_\r \bP^\s) + {\cal O}(k) \} )
\eeas
Similarly, $M^{+-ab \dot{A}}$ may be deduced from $J^{a \dot{A}}$ to be
\beas
M^{+-ab \dot{A}} &=& \str(e^{i k \cdot X} {i \sqrt{2} \over 4}
\sigma^{\dot{A}}_\ds {}^\dr \{  -2 D_a \bX^{\s \ds} D_b X_{\s \dr} 
+ i F_{ab} F^\ds {}_\dr  + {\cal O}(k) \} ) \; .
\eeas
Finally, the currents $T^{ab}$ and $J^{+-a}$ follows from $T^{+-}$ and $T_{\r 
\dr \; \s \ds}$ using the relations
\beas
T^{ab} &=& I^{ab} \; , \qquad J^{+-a} = I^{0a} \; , \qquad T_{\r \dr \; \s \ds} 
= I_{\r \dr \; \s \ds}\\
T^{+-} &=& - I^{00} - {1 \over 4} I^\mu {}_\mu - {1 \over 4} I^{\r \dr} 
{}_{\r \dr} 
\eeas
which again follow from Lorentz invariance of the dual Dp-D(p+4) brane actions. 
Using these, we find
\beas
T^{ab} &=& \str( e^{i k \cdot X} \{ F_a {}^\mu F_{\mu b} - D_a \bX^{\r \dr} D_b 
X_{\r \dr} - D_b \P_\r D_a \bP^\r - D_a \P_\r D_b \bP^\r \\
&& + b \delta^{ab} (-4 D_\mu \P_\r D^\mu \bP^\r + 4 F^\b {}_\a
\P_\b \bP^\a - 8 (\P_\a \bP^\s \P_\s \bP^\a - {1 \over 2} \P_\a \bP^\a \P_\s
\bP^\s) \} )\\
J^{+-a} &=& \str( e^{i k \cdot X} \{ F_{0b} F_{ab} + D_0 \bX^{\r \dr} D_a X_{\r 
\dr} + D_0 \P_\r D_a \bP^\r + D_a \P_\r D_0 \bP^\r
\eeas

We have now determined all terms in the matrix model currents that follow from 
the order $k^2$ terms in $T^{++}$. We have found that for consistency, the 
undetermined coefficients in $T^{++}$ must satisfy $a - b = - {1 \over 4}$. 
Using this relation to write $a$ in terms of $b$, and defining $\b = \a + 
\sqrt{2} b$ it is not hard to show that the complete set of terms in the 
effective action involving the undetermined coefficients $b$ and $\b$ may be 
written as
\be
\label{curvature}
S = \int dt d^9k \; {\cal R}_{++} \{2b \tr(\P_\r \bP^\r) \} + i
(\Theta^-_-)^\r \{ \b i \tr(\e_{\s \r} \c \bP^\s - \P_\r \chi^c) \}
\ee
where ${\cal R_{IJ}}$ is the Ricci tensor formed from the metric $h$ and 
$\Theta^I$ is the covariant field strength of the gravitino field,
\[
\Theta^I = \Gamma^{IJK} \partial_J \psi_K \; .
\]
The equations of motion for the graviton and gravitino in linearized 
supergravity are ${\cal R}_{IJ} = \Theta^I = 0$, so for any on-shell background 
fields, the terms with undetermined coefficients will vanish. Thus, we have 
uniquely determined the matrix theory operators coupling to a general set of 
on-shell supergravity fields (to the order at which we have worked). 

To define off-shell expressions for the currents, we will now argue that a 
natural choice is to take 
\[
b = \b = 0
\]
in the expressions for the currents 
above. Firstly, from (\ref{dim2currents}), we see that it is only for $b=0$ that 
we can write 
$T^{+a}$ and $J^{+ab}$ as $(0a)$ and $(ab)$ components of a covariant expression 
$I^{ab}$ as we had for the BFSS currents (the terms proportional to $k^a$ in 
$T^{+a}$ have no counterpart in $J^{+ab}$). Furthermore, from (\ref{auxil1}) and 
(\ref{auxil2}) we find that $b = \b = 0$ is the unique choice such that there 
are no auxiliary currents 
with purely bosonic terms (as we had for the BFSS theory). Finally, as 
we will see in sections 8 and 9, for $b=0$ the zero-brane charge 
density reproduces the expected distribution for simple configurations of 
instantons in D4-branes. Thus, to write final off-shell expressions for the 
currents we write all the undetermined coefficients in terms of $b$ and $\beta$ 
and assume that terms depending on $b$ and $\b$ belong in higher-order currents 
coupling to curvatures as in (\ref{curvature}) rather than in the basic currents 
defined above (the simplest possibility would be that $b=\b=0$). Our final 
results are summarized in the next section.
  
\section{Summary of Results for Dp-D(p+4) effective actions}

In this section, we summarize our results for the currents in Berkooz-Douglas 
model and use these to write down leading terms in the effective actions for all 
Dp-D(p+4) systems. Since these actions are all related by T-duality, which acts 
on worldvolume operators by dimensional reduction/oxidation, we will find it 
most convenient to write everything using the language of the D5-D9 system, 
which has the largest symmetry group. In particular, all the currents will be 
conveniently written in terms of a set of d=6 Lorentz covariant expressions 
which we now define.\footnote{Here, indices $\{\mu, \nu, \l, \kappa\}$ denote 6d 
Lorentz indeces while the remaining Greek indices are SU(2) indices, as usual.}

\noindent
At dimension 0, we define
\[
I_0 = \str(e^{i k \cdot X} \{1 - {1 \over 4} k_{\r \dr} \bar{k}^{\r \dr}
\P_\s \bP^\s + {\cal O}(k^3) \} )
\]
At dimension 2, we define
\beas
I_2^{\mu \nu} &=& \str ( e^{ik \cdot X} \{ 
-F^{\mu \nu } + {1 \over 8} \bl^\r \gamma^{\mu \nu a} \l_\r k_a + {1 \over 8}
\bt^\dr \gamma^{\mu \nu a} \t_\dr k_a + {\sqrt{2} \over 4} i \bar{k}^{\s
\ds} \e_{\s \r} \bl^\r \gamma^{\mu \nu } \t_\ds \\
&& \hspace{1in} + { 1 \over 4} \bc^c \gamma^{\mu \nu a} \c^c k_a  + {\cal O} 
(k^2) 
\} )\\
I_2^\mu {}_{\r \dr} &=& \str ( e^{ik \cdot X} \{ 
-D^\mu X_{\r \dr} - {\sqrt{2} \over 4} i \e_{\r \b} \bl^{\b}
\gamma^{\mu a} \t_\dr k_a - {1 \over 4} \bt^\ds \gamma^\mu \t_\dr k_{\r \ds}
- {1 \over 4} \bl^\s \gamma^\mu \l_\r k_{\s \dr}  \\
&& \hspace{1in}  - {i \over 2} k_{\b \dr} D^\mu \P_\r \bP^\b 
- {i \over 2} k_{\r \dr} \P_\s D^\mu \bP^\s + {i \over 2} k_{\b \dr} \P_\r D^\mu 
\bP^\b + {\cal O} (k^2)  \} )\\
I_2^A &=& \str ( e^{ik \cdot X} \{
{ \sqrt{2} \over 4} F^\a {}_\b \sigma^A_\a {}^\b - {\sqrt{2} \over 2}
\sigma^{A}_\r {}^\s \P_\s \bP^\r + {i \sqrt{2} \over
8} \bl^\r \sigma^A_\r {}^\s \gamma^a \l_\s k_a\\
&& \hspace{1in} + {1 \over 4} \bar{k}^{\r \dr}
\sigma^A_\r {}^\s \e_{\s \tau} \bl^\tau \t_\dr  + {\cal O} (k^2) \} )\\
I_2^{\dot{A}} &=& \str ( e^{ik \cdot X} \{ 
{ \sqrt{2} \over 4} F^\da {}_\db \sigma^{\dot{A}}_\da {}^\db 
+ {i \sqrt{2} \over
8} \bt^\dr \sigma^{\dot{A}}_\dr {}^\ds \gamma^a \l_\ds k_a - {1 \over 4} 
\e_{\ds \da} \bar{k}^{\a \dr}
\sigma^{\dot{A}}_\dr {}^\ds \bt^\da \l_\a + {\cal O} (k^2) \} ) \; .
\eeas
Finally, at dimension 4, we define
\beas
I_4^{\mu \nu} &=& \str( e^{i k \cdot X} \{ F^{\mu \a} F_{\a} {}^\nu - D^\mu 
\bX^{\r 
\dr} D^\nu X_{\r \dr} - D^\nu \P_\r D^\mu \bP^\r - D^\mu \P_\r D^\nu \bP^\r + 
{\cal O}(k) \} )\\
I_4^\mu {}_{\r \dr} &=& \str(e^{i k \cdot X} \{ F^{\mu \nu} D_\nu X_{\r \dr}
+ {i \over 2} D^\mu X_{\b \dr} F^\b {}_\r + {i \over 2} D^\mu X_{\r \db}
F^\db {}_\dr\\
& & \hspace{0.5in} - i D^\mu X_{\a \dr} \P_\r \bP^\a +
 {i \over 2} D^\mu X_{\r \dr} \P_\a \bP^\a + {\cal O}(k) \} )\\
(I_4)_{\r \dr \; \s \ds} &=& \str(e^{i k \cdot X} \{- {1 \over 2} 
\e_{\r \s} \e_{\dr
\ds} D^\mu \P_\a D_\mu \bP^\a - D_\mu X_{\r \dr} D^\mu X_{\s \ds}  
- { 1 \over 2} \e_{\s \a} \e_{\ds \da} F^\a {}_\r
F^\da {}_\dr  \\
&& \hspace{0.5in} 
- {1 \over 8} \e_{\r \s}
\e_{\dr \ds} (F^\a {}_\b F^\b {}_\a + F^\da {}_\db F^\db {}_\da ) 
+ \e_{\r \s} \e_{\dr \ds} F^\b {}_\a \P_\b \bP^\a
+ {1 \over 2} \e_{\r \a} \e_{\dr \db} F^\db {}_\ds \P_\s \bP^\a \\
&& \hspace{0.5in} + {1 \over 2} \e_{\s \a} \e_{\ds \db} F^\db {}_\dr \P_\r 
\bP^\a 
 - \e_{\r \s} \e_{\dr \ds} (\P_\a \bP^\s \P_\s \bP^\a 
- {1 \over 2} \P_\b \bP^\b \P_\s \bP^\s) + {\cal O}(k) \} )\\
I_4^{\mu \nu \l \kappa} &=& \str(e^{i k \cdot X} \{3 F^{[\mu \nu} F^{\lambda 
\kappa]} + {\cal O}(k) \} )\\
I_4^{\mu \nu \lambda} {}_{\r \dr}  &=& \str(e^{i k \cdot X} \{3 F^{[\mu \nu} 
D^{\lambda]} X_{\r \dr}  + {\cal O}(k) \} )\\
I_4^{\mu \nu A} &=& \str(e^{i k \cdot X} {i \sqrt{2} \over 4} \sigma^A_\s
{}^\r\{  (-2 D^\mu \bX^{\s \ds} D^\nu X_{\r \ds} + i F^{\mu \nu} F^\s {}_\r \\
&& \hspace{0.5in} - 2 (D^\nu \P_\r D^\mu \bP^\s - D^\mu \P_\r D^\nu \bP^\s  + i 
\P_\r \bP^\s
F^{\mu \nu}) + {\cal O}(k) \} )\\
I_4^{\mu \nu \dot{A}} &=& \str(e^{i k \cdot X} {i \sqrt{2} \over 4}
\sigma^{\dot{A}}_\ds {}^\dr \{  -2 D^\mu \bX^{\s \ds} D^\nu X_{\s \dr} 
+ i F^{\mu \nu} F^\ds {}_\dr  + {\cal O}(k) \} )\\
\tilde{I}_4^\mu {}_{\r \dr}  &=& \str(e^{i k \cdot X} \{   {i \over 6}
(D^\mu X_{\s \dr} F^\s {}_\r - D^\mu X_{\r \ds} F^\ds {}_\dr) 
-i D^\mu X_{\a \dr} \P_\r \bP^\a \\
&& \hspace{0.5in} + {i \over 2} D^\mu X_{\r \dr} \P_\a \bP^\a 
+ {\cal O}(k) \} )\\
\tilde{I}_4 &=& \str( e^{i k \cdot X} \{ {1 \over 8} F^\s {}_\r F^\r {}_\s 
- {1 \over 8} F^\dr {}_\ds F^\ds {}_\dr + {\cal O} (k) \} ) 
\eeas
Here, we have omitted the fermion terms in the dimension 4 currents, thought it 
would be straightforward to calculate these using the results of the previous 
section. 

\subsection{Results for Berkooz-Douglas matrix model currents}

In terms of the expressions we have just defined, the linear couplings of the 
eleven-dimensional supergravity fields to the Berkooz-Douglas matrix model are 
given by
\be 
\label{mtact3}
S_{MT} = \int dt d^9 k \; {1 \over 2} h_{IJ} T^{IJ} + { 1 \over 3!} 
A_{IJK} J^{IJK} + { 1 
\over 6!} A^D_{IJKLMN} M^{IJKLMN} + i S^{I} \psi_I \; . 
\ee
\\
\noindent
where\\

\be
\ba{lll}
T^{++} = I_0 & & \\
T^{+a} = I_2^{0a} & J^{+ab} = I_2^{ab}\\
T^+ {}_{\r \dr} = I_2^0 {}_{\r \dr} & J^{+a} {}_{\r \dr} = I_2^a {}_{\r \dr}\\
J^{+A} = I_2^A & J^{+ \dot{A}} = I_2^{\dot{A}} &\\
T^{+-} = - I_4^{00} - {1 \over 4} I_4^\mu {}_\mu - {1 \over 4} I_4^{\r \dr} 
{}_{\r \dr} \\
T^{ab} = I_4^{ab} & J^{+-a} = I_4^{0a} \\
T^a {}_{\r \dr} = I_4^a {}_{\r \dr} & J^{+-} {}_{\r \dr} = I_4^0 {}_{\r \dr} &\\
T_{\r \dr \; \s \ds} = (I_4)_{\r \dr \; \s \ds} & M^{+-} = \tilde{I}_4 &\\
M^{+-abcd} = I_4^{abcd} & J^{abc} = I_4^{0abc} &\\
M^{+-abc} {}_{\r \dr} = I_4^{abc} {}_{\r \dr} & J^{ab} {}_{\r \dr} = I_4^{0ab} 
{}_{\r \dr} & \\
M^{+-abA} = I_4^{abA} & J^{aA} = I_4^{0aA} & \\
M^{+-a} {}_{\r \dr} = \tilde{I}_4^a {}_{\r \dr} & J_{\r \dr} =
\tilde{I}_4^0 {}_{\r \dr} & 
\ea 
\ee

\noindent
Expressions for the fermionic currents $(S^+_+)_\r$ and $(S^+_+)_\dr$ with 
dimension ${3 \over 2}$ and $(S^+_-)_\r$, $(S^+_-)_\dr$, $(S_+^a)_\r$, 
$(S^a_+)_\dr$, $(S_{+ \r \dr})_\s$, and $(S_{+ \r \dr})_\ds$ with dimension   
${5 \over 2}$ may be found in the previous section. The remaining currents, all 
with dimensions greater than four, could be determined with additional work 
using the methods of the previous section.
 
\subsection{Results for Dp-D(p+4) brane actions}

The matrix theory currents of the previous subsection are related to leading 
terms in the currents for the D0-D4 system in type IIA string theory through the 
expressions (\ref{relationship}). Using these and the T-duality relations 
(\ref{tduality}), it is then 
straightforward to determine leading terms in the effective actions describing 
the linear couplings of type IIA/IIB supergravity fields to all Dp-D(p+4) brane 
systems. 

To describe the action, we take indices $\{\hat{I}, \hat{J}, \dots \}$ to run 
from 0 to 9, $\{ \hat{\mu}, \hat{\nu}, \dots \}$ to run from 0 to $p$ (the 
worldvolume directions of the Dp-brane), and $\{ \hat{a}, \hat{b}, \dots \}$ to 
run from $p+1$ to 5 (which we assume to be the directions transverse to the 
Dp-D(p+4) system). Finally, the D(p+4)-brane directions not shared by the Dp-
brane 
(which we assume to be 6,7,8,9) are denoted either by $SO(4)$ indices $\{i,j, 
\dots\}$ or by $SU(2) \times SU(2)$ indices $(\r \dr), (\s \ds), (\a \da), 
\dots$. Then the linear couplings to NS-NS fields are given by  
\be
\label{NScurrents}
S^{NS-NS} = \int d^{p+1} x d^{9-p} k \; {1 \over 2} h_{\hat{I} \hat{J}} 
{\cal T}^{\hat{I} \hat{J}} + \phi  {\cal J}_\phi  + 
{1 \over 2} B_{\hat{I} \hat{J}} {\cal J}_s^{\hat{I} \hat{J}}  
\ee
where 
\beas
&&{\cal J}_\phi = I_0 - {1 \over 4} (I_4^{\hat{\mu}} {}_{\hat{\mu}} + 
I_4^{\hat{a} \hat{a}} + I_4^{\r \dr} {}_{\r \dr} )\\
&&{\cal T}^{\hat{\mu} \hat{\nu}} = -\eta^{\mu \nu} I_0 - 
I_4^{\hat{\mu} \hat{\nu}} + {1 \over 4} \eta^{\hat{\mu} \hat{\nu}} 
(I_4^{\hat{\lambda}} {}_{\hat{\l}} + I_4^{\hat{a} \hat{a}} 
+ I_4^{\r \dr} {}_{\r \dr} ) \\
&&\ba{ll}
{\cal T}^{\hat{\mu} \hat{a}} = I_2^{\hat{\mu} \hat{a}} \qquad \qquad&
{\cal T}^{\hat{\mu}} {}_{\r \dr} =  I_2^{\hat{\mu}} {}_{\r \dr} \\ 
{\cal T}^{\hat{a} \hat{b}} = I_4^{\hat{a} \hat{b}}&
{\cal T}^{\hat{a}} {}_{\r \dr} = I_4^{\hat{a}} {}_{\r \dr}\\ 
{\cal T}_{\r \dr \; \s \ds} = (I_4)_{\r \dr \; \s \ds}&
{\cal J}_s^{\hat{\mu} \hat{\nu}} = -I_2^{\hat{\mu} \hat{\nu}}\\ 
{\cal J}_s^{\hat{\mu} \hat{a}} = I_4^{\hat{\mu} \hat{a}}&
{\cal J}_s^{\hat{\mu}} {}_{\r \dr} = I_4^{\hat{\mu}} {}_{\r \dr} \\ 
{\cal J}_s^{\hat{a} \hat{b}} = I_2^{\hat{a} \hat{b}}&
{\cal J}_s^{\hat{a}} {}_{\r \dr} = I_2^{\hat{a}} {}_{\r \dr}\\ 
{\cal J}_s^A = (I_2)_{A} & {\cal J}_s^{\dot{A}} = (I_2)_{\dot{A}} 
\ea
\eeas
To write the Ramond-Ramond couplings, it will be simplest just to write the
result for the D0-D4 action. The RR couplings for the other Dp-D(p+4)
systems may be obtained easily using the T-duality relations (\ref{tduality}). 
In
terms of the $I$ currents defined above, we have
\bea
S^{RR}_{D0-D4} &=& \int dt d^9 k \; C^{(1)}_0 I_0 \nonumber \\
&& + C^{(1)}_a I_2^{0a} + C^{(1) \; \r \dr} I_2^0 {}_{\r \dr} \nonumber \\
&& + {1 \over 2} C^{(3)}_{0ab} I_2^{ab} + C^{(3)}_{0a} {}^{\r \dr} 
I_2^a {}_{\r \dr} + C^{(3)}_{0A} I^A + C^{(3)}_{0 \dot{A}} I^{\dot{A}} 
\nonumber \\
&& + {1 \over 3!} C^{(3)}_{abc} I_4^{0abc} + {1 \over 2} C^{(3)}_{ab}
{}^{\r \dr} I_4^{0ab} {}_{\r \dr} + C^{(3)}_{aA} I_4^{0aA} +
C^{(3)}_{a \dot{A}} I_4^{0a \dot{A}} + C^{(3) \; \r \dr} 
\tilde{I}_4^0 {}_{\r \dr} \label{RRcurrents} \\
&& + {1 \over 4!} C^{(5)}_{0abcd} I_4^{abcd} + {1 \over 3!}  
C^{(5)}_{0abc} {}^{\r \dr} I_4^{abc} {}_{\r \dr} + {1 \over 2}
C^{(5)}_{0abA} I_4^{abA} + {1 \over 2} C^{(5)}_{0ab \dot{A}} I_4^{ab \dot{A}} +
C^{(5)}_{0a} {}^{\r \dr} \tilde{I}_4^{a} {}_{\r \dr} +  C^{(5)}_{0}
\tilde{I}_4 \nonumber \\
&& + \dots   \nonumber 
\eea
Here the dots indicate couplings to higher Ramond-Ramond fields 
which involve currents with dimensions six and higher.

\section{Open Dielectric branes}

In this section, we will use our results for the Dp-D(p+4) brane actions to 
produce and study configurations in which the lower dimensional branes are blown 
up into a noncommutative open D(p+2)-brane ending on the D(p+4)-brane. We will 
phrase the discussion in terms of the D0-D4 system, but most of the 
configurations we describe can be T-dualized to the higher dimensional cases.

\subsection{Flat open membranes from noncommutative instantons}

The simplest configurations of open D2-branes are planar configurations, for 
which the interior of the D2-brane lies completely inside the D4-brane. For 
these configurations, the D0-brane matrices $X^a$ corresponding to the 
directions transverse to the D4-brane should be set to zero. The remaining 
potential (before adding any background fields) is
\be
\label{v0}
V_0 = {1 \over 2} \tr({\cal D}_A {\cal D}_A) 
\ee
where
\be
\label{da}
{\cal D}_A  = \sigma^A_\r {}^\s ({1 \over 2} [\bX^{\r \dr}, X_{\s \dr}] - \P_\s 
\bP^\r )
\ee
We would like to turn on background fields which allow a cluster of D0-branes 
inside the D4-brane to expand into an open D2-brane. For the spherical 
dielectric branes of Myers, the strategy was to turn on a background field that 
made it energetically favorable for the system of D0-branes to carry a D2-brane 
dipole moment. This D2-brane dipole moment is measured by the operator coupling 
to the spatial derivative of $C_{0ij}$, thus, the appropriate background field 
was a constant $F_{0ijk}$. The present case is even simpler in that the 
configurations we want to produce have a non-zero charge, namely the D2-brane 
area in a given plane. Thus, it should suffice to turn on a constant RR 
potential $C_{0ij}$ where $i$ and $j$ are chosen to be in the directions of the 
D4-brane. 

From (\ref{RRcurrents}), we see that only the self-dual part of a constant 
$C_{0ij}$ field couples to the worldvolume fields (since $I^{\dot{A}}(k=0) = 
0$), and the relevant term in the D0-D4 Lagrangian is 
\[
C_{0A} \tr( {\sqrt{2} \over 2}  {\cal D}_A ) 
\]
with ${\cal D}_A$ defined in (\ref{da}) (actually the $X$ term vanishes upon 
taking the trace). But from (\ref{NScurrents}), we see that ${\cal D}_A$ is also 
the operator 
coupling to a self-dual NS-NS two form. Thus turning on either constant $C_{0A}$ 
or constant $B_A$ will introduce a term associating negative energy to an 
appropriately oriented D2-brane area.\footnote{We expect that terms coupling to 
$B_A$ and $C_{0A}$ at higher orders in $\a'$ will be different.} 

The constant $B_A$ deformation is something quite familiar: it is exactly the 
supersymmetric deformation leading to a self-dual noncommutativity parameter for 
the D4-branes. With this modification, the potential becomes (up to a constant)
\[
V = V_0 - {\sqrt{2} \over 2} B_A \tr( {\cal D}_A) = {1 \over 2} \tr(({\cal D}_A 
- {\sqrt{2} \over 2} B_A)^2) \; ,
\]
thus static solutions will preserve supersymmetry and satisfy 
\be
\label{adhm2}
{\cal D}_A = {\sqrt{2} \over 2} B_A
\ee
which are precisely the noncommutative version of the ADHM equations. Taking the 
trace of this relation, we find
\be
\label{area}
\tr( {\cal D}_A) = {\sqrt{2} \over 2} N B_A
\ee
which asserts that a configuration of $N$ noncommutative instantons carries a 
finite D2-brane area proportional to $N$ and to the $B$ field. Thus, we started 
out trying to find noncommutative open D2-branes and have ended up with 
noncommutative instantons. Nevertheless, we will 
now see that there do exist particular solutions with large numbers of 
instantons corresponding to large open D2-branes. Indeed, it was argued 
previously by Berkooz in \cite{berkooz} that the noncommutative ADHM moduli 
space should be associated with configurations of rigid (fixed area) open 
membranes in the matrix model of the light-like noncommutative (0,2) theory. 
This matrix model arises from the low energy limit of the background we are 
considering, so the noncommutative open D2-branes we are looking for should be 
exactly the same configurations as the large rigid open membranes of Berkooz.
We will now provide direct evidence for Berkooz's picture using the current 
operators we have derived and see that the many-instanton moduli space does 
contain large open D2-brane configurations.

We focus on configurations with a single D4-brane in which all the instantons 
sit in a single plane defined to be one of the two planes along which the 
self-dual B-field lies. For definiteness, we choose $B_A = (0,0,- \sqrt{2}B)$ 
corresponding to a self-dual $B$ field in the $12$ and $34$ directions, and look 
for solutions where the D0-brane coordinates $X^3$ and $X^4$ are zero. In terms 
of the $SU(2) \times SU(2)$ language, this implies $X_{12} = X_{21} = 0$. We 
relabel the remaining coordinates as $A = X_{11} = \bar{X}^{22}$. The 
noncommutative ADHM equations (\ref{adhm2}) then become
\beas
[A, A^\dagger] + \P_1 \bP^1 - \P_2 \bP^2 &=& B\\
\P_1 \bP^2 &=& 0
\eeas
In order for the second equation to be satisfied, we must choose either $\P_1$ 
or $\P_2$ to vanish. Taking the trace of the first equation, we see that for 
$B>0$ we must have $\P_1 \ne 0$ for the left side to be positive. Thus, we set 
$\P_2=0$ and define $\P = \P_1$ so that the remaining equation is 
\be
\label{const}
[A, A^{\dagger}] + \P \bP = B
\ee
This equation is precisely the constraint equations that arises in a matrix 
regularized version of noncommutative Chern-Simons theory suggested by 
Polychronakos,\footnote{This observation was exploited previously in \cite{hs} 
to suggest a string theory realization of quantum Hall fluids.}
\[
{\cal L} = {i \over B} \tr(A^{\dagger} D_0 A) + {i \over B} \bP D_0 \P + 
\tr(a_0)
\]
where $a_0$ is a gauge field $D_0 A = \dot{A} + i[a_0, A]$ and $D_0 \P = 
\partial_0 \P + i a_0 \P$. As discussed in \cite{susskind, polychronakos, hv} 
this model provides a good description of a the states of an 
incompressible quantum Hall fluid composed of $N$ electrons in the lowest Landau 
level.\footnote{Recall that the quantum phase space of a system of $N$ electrons 
in the lowest Landau level is reduced to a two-dimensional space which can be 
identified with the coordinate space. Thus the system behaves as if each 
electron is assigned a unit of area and the quantum degeneracy pressure prevents 
these areas from overlapping, resulting in an incompressible fluid-like 
behavior.} Using knowledge of the quantum Hall fluid states, we can then obtain 
a more physical picture of planar noncommutative instanton configurations.

\begin{figure}
\label{puddles}
\centerline{\epsfysize=3truein \epsfbox{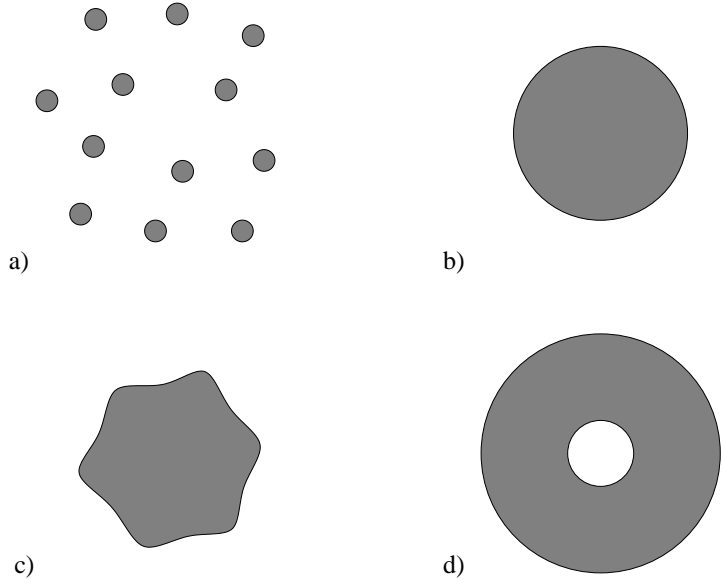}}
 \caption{Configurations of noninteracting electrons in the Lowest
 Landau level, or planar configurations of noncommutative instantons.}
\end{figure}

A variety of configurations of electrons in the lowest Landau level are depicted 
in figure 3. Generic states (a) have electrons which are widely 
separated, and 
clearly correspond to states with widely separated instantons. At the opposite 
extreme are states for which all of the electrons come together to form a 
``puddle'' of incompressible quantum Hall fluid. This puddle may be round (b) 
(corresponding to the ground state in the case that the electrons sit in a 
harmonic oscillator potential) or any other shape (c),(d) as long as the total 
area is $BN$.\footnote{Note that the shapes are ``fuzzy'' just as for closed 
membrane states in a matrix approximation. Precise shapes will be recovered only 
in the limit of large $N$.} It is noncommutative instanton configurations 
corresponding to these puddle states that we would like to identify with large 
open membranes. 

In the next subsections, we will write down explicit solutions of (\ref{const}) 
corresponding to some of these puddle states (borrowed from Polychronakos' 
studies of quantum Hall states) and show that may indeed be interpreted as open 
D2-branes.

\subsection{The open D2-brane disk}

The simplest ``puddle'' solution to (\ref{const}) is that corresponding to the 
round quantum Hall droplet (figure 3b), and is given by 
\be
\label{disksol}
A = \sqrt{B} \left( \ba{ccccc} 
 0 & 1 & & & \\ 
 & 0 & \sqrt{2} & & \\
 & & & \ddots & \\
 & & & 0 & \sqrt{N-1}\\
 & & & & 0
\ea \right) 
\qquad \qquad
\P = \sqrt{B} \left( \ba{c} 0 \\ 0 \\ \vdots \\ 0 \\ \sqrt{N} \ea \right)
\ee
This was written down by Braden and Nekrasov \cite{bn} in the context of 
noncommutative instantons and by Polychronakos \cite{polychronakos} to describe 
the ground state of electrons in a harmonic oscillator potential in the lowest 
Landau level. It was argued by Berkooz in the context of the lightlike 
noncommutative (0,2) theory that this configuration corresponds to a round rigid 
open membrane. 

\begin{figure}
\centerline{\epsfysize=2truein \epsfbox{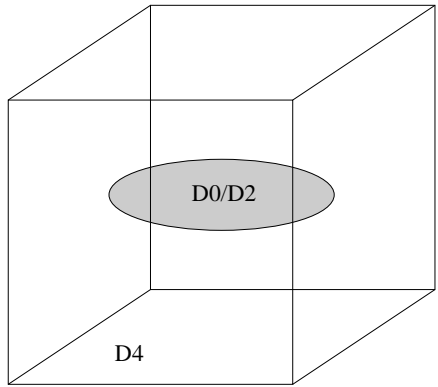}}
 \caption{Noncommutative D2-brane disk in a D4-brane.}
\end{figure}

There are various ways to see that this indeed corresponds to a round
open D2-brane. 
First, equation (\ref{area}) shows that this configuration has a net D2-brane 
area in the $12$ plane given by $2 \pi B N$, since the membrane charge 
(operator coupling to $C_{012}$) is $\tr({\cal D}_3)$ and the membrane tension 
in our units is $2 \pi$. This however is true of all the noncommutative 
instanton configurations since each instanton carries a unit of area. To measure 
how this area is distributed, it is useful to determine the moments of the 
charge distribution. From (\ref{RRcurrents}), moments of the zero-brane charge 
distribution are 
given by derivatives of ${\cal J}^{(1)}_0 = T^{++}$ at $k=0$. The dipole moment 
is
\[
-i \partial_{k_i} {\cal J}^{(1)}_0 (k=0) = \tr(X_i) = 0
\]
so the charge distribution is centered at the origin. The quadrupole moments in 
the $x-y$ plane are
\[
I_{ij} = \partial_{k_i} \partial_{k_j} {\cal J}^{(1)}_0 (k=0) = \tr(X_i X_j + {1 
\over 
2} \delta_{ij} \P \bP )
\]
It is straightforward to calculate 
\[
I_{xy} = 0, \; I_{xx} = I_{yy} = {1 \over 2} B N^2
\]
These are exactly the quadrupole moments of a distribution of $N$ units of 
charge spread uniformly over a round disk of area $2 \pi BN$. Further, the 
moment of inertia for the D0-brane charge, $I_{xx} + I_{yy}$ is given by the 
trace of the operator
\be
\label{rsquared}
r^2 \equiv X^2 +Y^2 + \P \bP = B \left( \ba{ccccc} 
 1   & & & \\ 
 & 3   & & \\
 & &  \ddots   & \\
 & & & 2N-1
\ea \right) \; .
\ee
The fact that this operator is diagonal with evenly spaced eigenvalues suggests 
that the area is evenly spaced with respect to the radius squared up to a 
maximum $r^2$ of $\sim 2BN$, as we would expect for a uniform disk. Finally, we 
note  that in the $N = \infty$ limit, the matrix $A$ goes over to the matrix 
representation of a harmonic oscillator creation operator, satisfying $[A, 
A^{\dagger}] = B$. In terms of the real coordinates, this is $[X,Y] = iB$, which 
is exactly the configuration of an infinite number of D0-branes describing an 
infinite flat noncommutative D2-brane.  

Together, these observations give good evidence that the noncommutative 
instanton configuration (\ref{disksol}) may indeed be interpreted as an open 
noncommutative D2-brane disk of radius $R = \sqrt{2BN}$. It also provides an 
example of how the current operators we have derived may be used to learn about 
the spacetime distribution of charges for a given configuration.

\subsection{Other planar open D2-branes}

There are a number of transformations that permit us to generate new solutions 
of (\ref{const}) starting from any given solution.

As pointed out in \cite{polychronakos}, given a solution of (\ref{const}), the 
infinitesimal transformation
\be
\label{inf1}
A \to A + \epsilon_n (A^\dagger)^n
\ee
preserves the condition (\ref{const}) and therefore generates a new solution. As 
explained in \cite{polychronakos}, these correspond to area-preserving 
deformations of the disk which change the shape of the boundary of the membrane 
(starting from the disk, they produce ripples with wavelength $\propto {1 \over 
n}$). They are related to infinitesimal area preserving diffeomorphisms of the 
complex plane 
\[
z \to z + h
\]
given by 
\be
\label{z1}
h = \e_n \bar{z}^n
\ee
In fact, any function $h$ satisfying $\partial h + \bar{\partial} \bar{h} = 0$ 
will generate an area-preserving diffeomorphism, and in general we may choose $h 
= \bar{\partial} (g - \bar{g})$. These are generated by (\ref{z1}) as well as 
\be
\label{z2}
h = \e m z^n \bar{z}^{m-1} - \bar{\e} n \bar{z}^{n-1} z^m 
\ee
From this latter set of generators, we may guess another set of solution 
generating infinitesimal transformations for $A$, namely
\be
\label{inf2}
A \to A + \e_{mn} {\rm sym} (A^n (A^\dagger)^{m-1}) - \bar{\e}_{mn} {\rm sym}
(A^m (A^\dagger)^{n-1})
\ee
where ${\rm sym}$ indicates a summing over all possible orderings with 
coefficient 1 for each independent term.\footnote{The number of terms in the 
first expression is ${m \over n}$ times the number of terms in the second, so we 
should not include the coefficients $m$ and $n$ that appeared in (\ref{z2}).} It 
is not hard to show that these transformations preserve the constraint 
(\ref{const}) to leading order in $\e$. 

Not all of these transformations are independent, firstly because $A^{N+1}$ may 
be written in terms of lower powers of $A$ and also because some of these may be 
equivalent to infinitesimal gauge transformations
\[
A \to A + i[B + B^\dagger, A] \qquad \P \to \P + i (B + B^\dagger) \P \; .
\]

It is straightforward to integrate some of the simpler transformations to 
explicitly produce new solutions. The constant $h$ transformations clearly 
generate translations $A \to A + z \identity$, while those linear in $z$ give 
$A \to \alpha A + \beta A^\dagger$ with $|\a|^2 - |\b|^2 = 1$, corresponding to 
$SL(2,R)$ transformations on the plane (rotations, shears, squashings). For 
example, starting from the disk and performing the $SL(2,R)$ transformation that 
expands the $x$ direction while contracting the $y$ direction gives the solution 
\[
A = \sqrt{B} \cosh(a) \left( \ba{ccccc} 
 0 & 1 & & & \\ 
 & 0 & \sqrt{2} & & \\
 & & & \ddots & \\
 & & & 0 & \sqrt{N-1}\\
 & & & & 0
\ea \right) + \sqrt{B} \sinh(a) \left( \ba{ccccc} 
 0 &  & & & \\ 
 1 & 0 & & & \\
 & \sqrt{2} & 0 &  & \\
 & & & \ddots & \\
 & & & \sqrt{N-1} & 0
\ea \right) 
\]
which should correspond to an ellipsoidal membrane with axes along the 
coordinate axes. As for the disk, one may calculate moments of the charge 
distribution, and one finds agreement with this geometrical picture.

A somewhat more complicated transformation that can be integrated readily for 
the case of the disk is 
\[
A \to A + \epsilon (A^\dagger)^{N-1} \; .
\]
As a transformation on the complex plane, this introduces $n$ ripples on the 
boundary of the unit disk, however, we will find a rather different 
interpretation for the transformation on our noncommutative open membrane disk. 
Solving
\[
\partial_t A = (A^\dagger)^{N-1} \; ,
\]
one finds the solution
\[
A = \sqrt{B}\left( \ba{ccccc} 
 0 & \sqrt{1 + q} & & & \\ 
&  0   & \sqrt{2 + q} & & \\
&   & 0  & & \\
&   & & \ddots  & \sqrt{N-1 + q} \\
\sqrt{q} & & &   & 0
\ea \right) 
\qquad \qquad
\P_1 = \sqrt{B} \left( \ba{c} 0 \\ 0 \\ \vdots \\ 0 \\ \sqrt{N} \ea \right)
\]
where $q$ is an arbitrary parameter.
This is exactly the solution written down by Polychronakos to describe a quantum 
Hall droplet with a quasihole of charge proportional to $q$ at the center. For a 
large quasi-hole, the electron fluid forms an annulus, so it seems reasonable to 
identify this solution with an annular open D2-brane. To verify this, note that 
the moment of inertia matrix for this solution is
\[
r^2 = A A^\dagger + A^\dagger A + \P \bP  = B \left( \ba{ccccc} 
 1 + q & &  & & \\ 
 & 3 + q &  & & \\
 & &  \ddots  & \\
 & & & 2N-1 + q &
\ea \right) \; .
\]
As for the disk, we see that the elements of area are equally spaced in $r^2$, 
but this time from $r^2 \sim Bq$ to $r^2 \sim (2N + q)b$, consistent with the 
interpretation as a uniform annulus with inner and outer radii $\sqrt{qB}$ and 
$\sqrt{(2N + q)B}$ and total area $2 \pi B N$, as before.

\begin{figure}
\centerline{\epsfysize=2truein \epsfbox{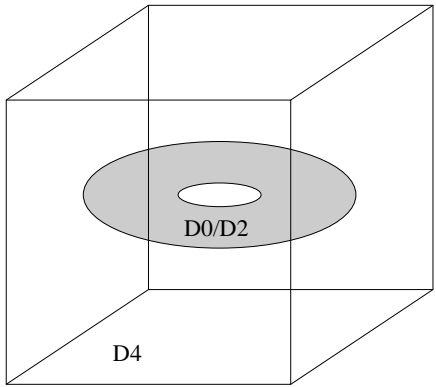}}
 \caption{Annular open D2-brane configuration.}
\end{figure}

Thus, area preserving diffeomorphisms of the complex plane map to a set of 
transformations (\ref{inf1}) and (\ref{inf2}) that include both ``smooth'' 
boundary deformations 
of a D2-brane disk as well as transformations which create a new boundary. Using 
this set of transformations, it should be possibly to produce solutions 
corresponding to planar open membranes of arbitrary shape and topology (in the 
limit of large $N$). T-dualizing in directions transverse to the D4-brane, we 
can produce an analogous set of solutions corresponding to open D(p+2)-branes 
ending on D(p+4)-branes. In particular, for $p > 1$, there will be a real moduli 
space, and these solutions will correspond to different supersymmetric vacua.

\subsection{Bulging branes}

We have seen in the previous section that there exist planar configurations of 
noncommutative instantons which have charge distributions corresponding to open 
D2-branes of various shapes. For these configurations, the bulk of the D2-brane 
sat completely inside the D4-brane, and all of the configurations were 
degenerate. In particular, the open D2-brane configurations could fall apart at 
no cost in energy to a collection of widely separated D0-branes dissolved in the 
D4-brane, each carrying a unit of membrane charge. 

It is reasonable to ask whether such an open D2-brane configuration really 
behaves like a large membrane as opposed to simply a collection of individual 
instantons each carrying some membrane charge and arranged in the shape of a 
large membrane. One way to address this question would be to try and excite some 
collective mode of the membrane. In particular, we will attempt to pull the 
interior of our D2-brane disk off the D4-brane by turning on additional 
background fields. Since the membrane carries both D0-brane charge and D2-brane 
charge, we could exert a force on the membrane in a direction transverse to the 
D4-brane either by turning on a gradient of $C^{(1)}_0$ or $C^{(3)}_{012}$. We 
will 
choose to pull the membrane off by its D0-brane charge\footnote{The other choice 
gives a similar solution.}, so we turn on an 
additional weak background field given by
\[
C^{1}_{0} = \alpha x^9
\]
where $x^9$ is one of the directions perpendicular to the D4-brane. Including 
the new operator coupling to this background field (from (\ref{RRcurrents})) as 
well as the 
additional terms in the potential involving the matrix $X^9$, the potential 
becomes 
\bea
V &=& \tr( - [A , X^9][A^\dagger , X^9]) + \bP X^9 X^9 \P \nonumber \\
& & + {1 \over 2} \tr(([A, A^\dagger] + \P \bP - B)^2) \nonumber \\
& & + \alpha \tr( X^9 ) \label{potl}
\eea
We now start with the solution (\ref{disksol}) corresponding to the D2-brane 
disk and look 
for a new solution perturbatively in $\a$. The relevant equations of motion are
\beas
0 &=& [A, [A^\dagger, X^9]] + [A^\dagger, [A, X^9]] + \P \bP X^9 + X^9 \P \bP + 
\a\\
0 &=& [[A,X^9],X^9] + [[A,A^\dagger] + \P \bP , A] \\
0 &=& X^9 X^9 \P + ([A, A^\dagger] + \P \bP - B) \P
\eeas
We expect that the final configuration should retain the rotational symmetry in 
the 1-2 plane, so that each value of $r^2$ (where $r$ is the distance from the 
centre of the D2-brane) should correspond to a particular value of $X^9$. Since 
the $r^2$ matrix (\ref{rsquared}) of the unperturbed solution was diagonal, we 
therefore look 
for a solution for which $X^9$ is also diagonal. It is not hard to show that to 
leading order in $\a$, the equations of motion are solved by   
\[
A = A_0 + {\cal O}(\a^2) \; , \; \;  \P = \P_0 + {\cal O} (\a^2) \; , \; \; X^9 
= - {\alpha \over 2B} \left( \ba{ccccc} 
 N & &  &  \\ 
 & N-1 &  &  \\
 & &  \ddots  & \\
 & & & 1  \ea \right) +  {\cal O} (\a^2)
\]
where $A_0$ and $\P_0$ are the unperturbed disk solution given in 
(\ref{disksol}). Recalling 
that the $r^2$ matrix was
\[
r^2 = B \left( \ba{ccccc} 
 1  & &  &  \\ 
 & 3 &  &  \\
 & &  \ddots  & \\
 & & & 2N-1  
\ea \right)
\]
we see that 
\[
X^9 = - {\a \over 4B^2}(B (2N+1) - r^2)
\]
is satisfied exactly as a matrix equation. This suggests that the perturbed 
membrane solution has a well defined parabolic $x^9(r)$ profile given by
\[
x^9 \sim - {\a \over 4B^2} (R^2 - r^2) \; , 
\]
as depicted in figure 6. Thus, only the boundary of the D2-brane 
remains attached to the D4-brane. 

Substituting our solution back into the potential (\ref{potl}), we find that to 
order $\a^2$, the energy of the perturbed solution is 
\[
E = - { \a^2 \over 8 B^2} N (N+1) \approx - { \a^2 \over 32 B^4 } R^4 = - { \a^2 
\over 32 \pi^2 B^4} {\rm Area}^2  
\]
Since this energy scales like the square of area, it is clear that two widely 
separated D2-brane disks will prefer to come together to form a single bulging 
D2-brane rather than two separate ones. Similarly, we expect that upon turning 
on this background field, a general planar collection of noncommutative 
instantons will prefer to merge into a single open-D2 brane which allows for the 
maximal bulge off the D4-brane. It would be interesting to verify this more 
explicitly. 

\begin{figure}
\label{bulge}
\centerline{\epsfysize=1.25truein \epsfbox{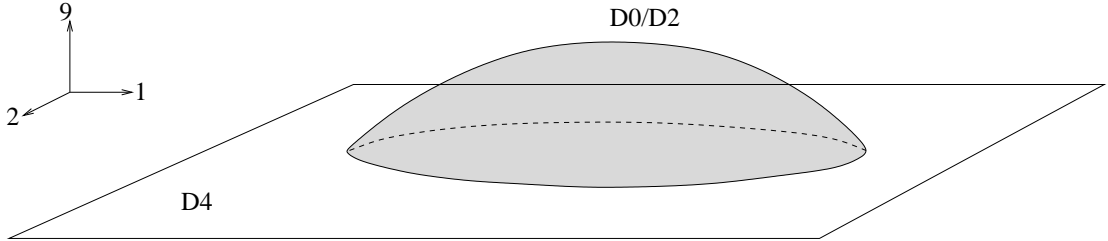}}
 \caption{Bulging noncommutative open D2-brane.}
\end{figure}

We should note that just as for the constant $F_{0ijk}$ in Myers' spherical 
D2-brane example, the constant $\partial_9 C^{(0)}$ field we have considered is 
not 
by itself a consistent background of string theory since it carries energy 
density and will induce a non-zero curvature. However, for weak fields, the 
effects from this curvature will be at higher order in the small parameter $\a$, 
so we do not expect the resulting configuration for a consistent string theory 
background to be qualitatively different.

\section{Higher order terms}

In section 6, we derived leading terms in the currents of the Berkooz-Douglas 
matrix model starting with a rather incomplete expression for $T^{++}$. We found 
that the leading terms in this expression involving the fundamental fields were 
actually fixed by demanding consistency with the supersymmetry relations 
(\ref{master}), various conservation relations, and symmetries. It is possible 
that higher order terms in $T^{++}$ might be determined in a similar way from 
these terms. Rather than following this approach and attempting to derive these 
higher order terms, in this section we use a different approach to suggest a 
more complete expression for $T^{++}$. If correct, this could be used as a 
starting point to derive complete expressions for the remaining currents in the 
Berkooz-Douglas model.

We begin by considering $N$ D0-branes in the presence of $k$ coincident 
D4-branes in the weak coupling, $\alpha' \to 0$ limit of type IIA string theory. 
In this context, our desired current $T^{++}$ is the operator which measures the 
density of D0-brane charge. Specifically, we would like to consider 
supersymmetric configurations on the (classical) Higgs branch of theory where 
all the D0-branes are dissolved in the D4-brane. These configurations have two 
alternate descriptions in string theory. 

In the first picture, we describe the configuration explicitly in terms of the 
0-0 and 0-4 string degrees of freedom.  Supersymmetric configurations are those 
for which the potential in the D0-brane quantum mechanics vanishes,
\be
\label{ADHM}
\sigma^A_\r {}^\s ({1 \over 2} [\bX^{\r \dr}, X_{\s \dr}] - \P_\s 
\bP^\r ) = 0 \; .
\ee
Note that we are assuming $X^a$ is zero since we are on the Higgs branch. For 
these solutions, it is consistent to set the D4-brane fields arising from 4-4 
stings to zero. 

In the second picture, we describe everything in terms of the gauge field on the 
D4-brane. Here, supersymmetric configurations are $N$-instanton solutions to the 
Yang-Mills equations. In this description, we simply do not include the 0-0 
string or 0-4 string degrees of freedom, since these are already described by 
the parameters describing the locations and sizes of the instantons.\footnote{In 
fact, there would seem to be many other descriptions for which we describe $m$ 
of the instantons explicitly using the 0-0 and 0-4 string degrees of freedom and 
take the gauge field on the D4-brane in the $(N-k)$-instanton sector.} For this 
description to be valid, the scale size of the instantons should be much larger 
than the string scale. 

The situation is similar to a set of D3-branes in string theory which may be 
described either by including the D3-brane worldvolume degrees of freedom 
explicitly and choosing a flat background for the closed strings or by thinking 
completely in terms of closed strings moving on the D3-brane geometry. 

For the D0-D4 case, there is an explicit map between the two descriptions given 
by the ADHM construction. This construction takes fields $X_{\r \dr}$ and 
$\P_\r$ 
satisfying the ADHM equations (\ref{ADHM}) and computes a gauge field 
$A_i (X_{\r \dr},\P_\r ;x^i)$ which is a self-dual solution to the Yang-Mills 
equations in the $N$ instanton sector. This construction will be very useful for 
our purposes, since in the pure D4-brane language, the operator measuring 
D0-brane charge is well known, 
\be
\label{fwedgef}
\r(x) = \tr(F \wedge F) \; .
\ee
Since the D0-brane charge density should be the same regardless of what 
description we use, it must be that the D0-brane density in the first picture is 
given by (\ref{fwedgef}) where $F$ is taken to be the field strength computed 
from the gauge field determined by the ADHM construction from $X_{\r \dr}$ and 
$\P_\r$.

We will now be more explicit. Let the spatial coordinates of the D4-brane be 
denoted $x^i$, or in the $SU(2) \times SU(2)$ notation, $x_{\r \dr}$. Then given 
fields $X_{\r \dr}$ and $\P_\r$ satisfying (\ref{ADHM}) and defining $Z_{\r \dr} 
= 
X_{\r \dr} - x_{\r \dr}$, the ADHM gauge field is given by
\[
A_i = \Lambda^{\dagger} \partial_i \Lambda + \psi_\dr^\dagger \partial_i 
\psi^\dr
\]
where $\Lambda$ and $\psi_\dr$ are each $k \times k$ matrices satisfying
\be
\label{aux}
Z_{\r \dr} \psi^\dr = \P_\r \Lambda 
\ee
and normalized such that 
\[
\Lambda^{\dagger} \Lambda + \psi_\dr^\dagger \psi^\dr = \identity
\]
For generic configurations, we may use these formulae to calculate the density 
$\tr(F \wedge F)$ directly in terms of $X$ and $\P$. Defining
\[
f = 2 (Z_{\r \dr} \bar{Z}^{\r \dr} + \P_\r \bP^\r)^{-1} 
\]
and
\[
B_\rho^{\sigma} = {1 \over 2 \pi} f (\delta^\s_\r - 
\bar{Z}^{\s \dr} f Z_{\r \dr}) 
\]
we find (for more details, see for example \cite{kms})
\bea
\r(x) &=& \tr(F \wedge F) \nonumber\\ 
&=& \tr((B_\r^\s \sigma^B_\s {}^\a)^2) \label{gent} \\
&=& \tr(2 B_\r^\r B_\s^\s - B_\r^\s B_\s^\r) \nonumber
\eea
This is the position space D0-brane charge distribution for a general 
configuration on the instanton moduli space.\footnote{The expression we have 
written is well defined for generic configurations of instantons, but is 
singular at certain points on the moduli space corresponding to small instanton 
singularities.} Since the ADHM mapping assumes that the ADHM equations are 
satisfied, the operator $\rho(x)$ can only be expected to agree with our desired 
operator $T^{++}$ up to terms that vanish when the equations of motion are 
satisfied. Thus, we may conclude that
\[
T^{++} (k^a = X^a = 0, {\rm on \; shell}) = \r(x)
\]
To make a comparison with our previous expression for $T^{++}$, we should go to 
momentum space, however it seems rather tricky to Fourier transform the general 
expression $\r(x)$. Things simplify considerably if we consider the special case 
of a single D0-brane. Here, the expression for 
$\r$ becomes simply 
\be
\label{falloff}
\r(x) = {6 \over \pi^2} { (\P_\r \bP^\r)^2 \over (\P_\r \bP^\r + |x-X|^2)^4}
\ee
After rewriting the denominator using 
\[
{1 \over t^4} = {1 \over 6} \int_0^\infty d \a \a^3 e^{-\a t}
\]
it is straightforward to transform to momentum space, and we find
\be
\label{compl}
\r(k) = e^{i k \cdot X} \int_0^\infty d \a \a e^{-\a - {1 \over 
4 \a} k^2 \P_\r \bP^\r }
\ee
Expanding in powers of momentum, we find
\[
\r(k) = e^{i k \cdot X} (1 - {1 \over 4} \bar{k}^{\r \dr} k_{\r \dr}  \P_\r 
\bP^\r + {\cal O}(k^4)) 
\]
which precisely agrees with our previous expression for ${\cal J}_{(1)}^0 = 
T^{++}$. This 
gives evidence that (\ref{compl}) may indeed provide a complete expression for 
$T^{++}$ for the case of a single D0-brane, up to terms that vanish when the 
ADHM equations 
\[
\P_\r \bP^\s = {1 \over 2} \delta_\r^\s \P_\a \bP^\a \; .
\]
are satisfied. The simplest possibility would be that these extra terms are zero 
and that (\ref{compl}) is the correct operator off-shell. To test this, one 
could use the methods of section 6 and check whether this expression is a 
consistent starting point for the supersymmetry relations.

To summarize, we have argued that a complete expression for the zero-brane 
density operator ${\cal J}_{(1)}^0 = T^{++}$ in the $\a' \to 0$ limit may be 
given in position space by (\ref{gent}) and in momentum space for the case of a 
single D0-brane by (\ref{compl}), up to terms which vanish when the ADHM 
equations are satisfied. Assuming the validity of these expressions, one could 
use the methods of section 6 to obtain more complete expressions for the 
remaining currents.

\subsection{A puzzle}

Up to order $k^2$, we found that the expression $\r(k)$ in (\ref{compl}) agrees 
with our 
previous expression for $T^{++}$. However, note that if one tries to expand the 
expression (\ref{compl}) further, one finds that the $k^4$ term is infinite (the 
$\a$ integral diverges once we bring down two powers of ${1 \over \a}$). Thus, 
while the expression (\ref{compl}) is clearly well defined for all values of 
$\P$ 
and $k$ (the integral is always finite), it is not analytic at $k=0$. The 
leading singularity is of the form $k^4 \log(k^2)$. 

In fact, this behavior should have been expected. Recall that various terms in 
the momentum expansion of the current $T^{++}$ correspond to multipole moments 
of the zero-brane charge distribution. However from (\ref{falloff}), it is clear 
that the charge distribution for an instanton configuration falls off only 
algebraically, as ${1 \over r^8}$ in the D4-brane directions. Thus, integrating 
this distribution over the D4-brane worldvolume with more than three powers of 
$x^i$ will give a divergent result. In other words, multipole moments beyond the 
octopole moment cannot be defined since the charge distribution does not fall 
off fast enough. Thus the non-analyticity of $T^{++}$ at $k=0$ is consistent 
with physical expectations. This is not the puzzle.

The puzzle comes when we consider the string theory effective action, which 
includes the term
\be
\label{stringea}
\int dt d^9 k \; C^{(1)}_0(-k,t) T^{++}(k,t) \; .
\ee
Typically, one expects to be able to calculate terms in the classical string 
theory effective action by computing tree-level amplitudes with specific numbers 
of vertex operators. For example, the $C^{(1)} \P \bP$ term should come from a 
disk amplitude with one Ramond-Ramond closed string vertex operator inserted on 
the bulk of the worldsheet and two $0-4$ open string vertex operators inserted 
on the boundary. On the other hand, the terms in (\ref{stringea}) with four 0-4 
fields are (by dimensional analysis) also the terms at order $k^4$ in a momentum 
expansion, and we have argued that such terms are not well defined on their own 
(since the 16-pole moment is not well-defined). This suggests that in computing 
the five-point function with one RR vertex operator and four 0-4 string vertex 
operators, one would encounter a divergence that could be cured only by summing 
over the complete set of disk amplitudes with arbitrary numbers of 0-4 vertex 
operators.\footnote{Note that the five-point function is expected to contain 
poles corresponding to tree level field theory diagrams with intermediate 
states. The $C \P \bP \P \bP$ term in the effective action (\ref{stringea}) is 
the 1PI contribution which remains after subtracting off the other tree-level 
field theory contributions. Our discussion suggests that even this
remaining term may have some divergence.} To reiterate, if the $C^{(1)} k^4
\P^4$ term by itself
were finite, it would indicate that instanton configurations should
have well defined 16-pole moments, inconsistent with the ${1 \over r^8}$ 
falloff. 

One way around this conclusion could be that the charge density
operator takes the singular form (\ref{compl}) only after integrating
out some light field. A natural candidate might be the 4-4 string fields,
but these can be included explicitly in the effective action and give
the usual $C_0 Tr(F \wedge F)$ term. In the description of instantons
in terms of the 0-0 and 0-4 string degrees of freedom, the D4-brane
gauge field can be consistently set to zero, so this additional term
does not contribute to the charge density. Thus, it is not clear to us
what the extra light field could be.

If it is true that the five point function has a divergence, we would
have the interesting situation of apparently having to sum over an
infinite set of string amplitudes to obtain the effective action. This
is reminiscent of dealing with infrared divergences in field
theory. Before speculating further, it seems worthwhile to directly study the
five point function in question to check whether there is actually a divergence.
We leave this as a problem for future work.

\section{Remarks}

In this paper, we have derived leading terms in the operators which
describe supergravity currents in the Berkooz-Douglas matrix model of
M-theory with M5-branes. We used these operators to write down leading
terms in the action describing linear couplings of type II
supergravity fields to systems with arbitrary numbers of Dp-branes and
D(p+4)-branes in string theory. Using these explicit actions, we
demonstrated situations in which turning on certain background fields
allows a collection of D0-branes to form a stable open D2-brane
ending on a D4 brane, including planar configurations in which the
D2-brane sits completely inside the D4-brane, as well as a bulging
brane configuration with the D2-brane attached only at the
boundary. The current operators we derived provide a valuable tool to
determine the spacetime configuration of charges for a given matrix
configuration and thus to provide a geometrical picture for the
configurations we have studied.

There are numerous directions for future work. Firstly, it would be
interesting to determine more complete expressions for the currents. 
We have explicitly calculated leading terms
in a momentum expansion (corresponding to low order moments) of the
currents starting with the leading terms in the primary current
$T^{++}$ (corresponding to the zero-brane density). A more complete
expression (up to equation of motion
terms) for $T^{++}$ was proposed in section 9, based on the ADHM
construction, and this could serve as a starting point for deriving full
expressions for the other currents, though the expression we have
given is only required to be valid on-shell. It is possible that some
of the ambiguous terms (possible extra terms which vanish when the
ADHM equations are satisfied) might be fixed by consistency with the
supersymmetry relations. If the expression we have proposed is
correct, it would also be interesting to understand how it could arise
from string theory amplitudes, since it does not have a well defined
expansion in momenta or in the number of open string fields.

It would also be interesting to find and study other noncommutative
open brane configurations using the actions we have derived. For
example, there should be situations (e.g. ones related to those we have studied
by boosting) for which the open D2-brane is stabilized by a
combination of background fields and motion within the D4-brane
(analogous to the closed brane configurations studied in \cite{dtv}). Other
interesting configurations might arise in situations with
more than one D4-brane, perhaps separated in some transverse
direction. More generally, it would be nice to have a description of
arbitrarily shaped open D2-branes for various topologies (such as the
one for closed membranes of spherical \cite{kt1} and toroidal \cite{bfss}
topology). The solution-generating set of transformations given in
section 8.3 would probably be quite useful in this regard.

We have seen that the actions for various Dp-D(p+4) systems are
related by T-duality in the directions transverse to both sets of
branes or shared by both sets of branes. One could also perform
T-duality in the other directions to derive actions for perpendicular
brane systems. For example, starting with the D0-D4 system and
T-dualizing in two of the D4-brane directions, we arrive at a system
with two perpendicular sets of D2-branes. Couplings of supergravity
fields to the worldvolume fields in this system would follow
immediately from the results of this paper.

Another application would be to the study of 3+1 dimensional field
theories living on D3-branes in the presence of D7-branes. These are
${\cal N} = 2$ supersymmetric gauge theories in the absence of
background fields, but can be deformed to theories with less
supersymmetry by turning on various operators. Our results
give all relevant and marginal operators that can be turned on in the
presence of any constant supergravity potential or field strength.

A rather different application of this work might be to the search for
an off-shell formulation of eleven-dimensional supergravity. 
In deriving the supersymmetry relations between Matrix theory
currents, we needed to include ``auxiliary'' terms to account for the
fact that the supersymmetry variation of the action is only
required to vanish on shell, that is when the equations of motion for
the bulk supergravity fields are satisfied. If an off shell formulation of
D=11 supergravity exists, one should be able to write down a set of
currents coupling to all the fields including the auxiliary
fields. This expanded set of currents would then be required to obey a
set of supersymmetry relations without any auxiliary terms. By
examining the structure of the auxiliary terms we have derived
(e.g. in the BFSS Matrix model) and interpreting these as a set of
auxiliary currents, one might be able to deduce some of the auxiliary
fields that would be required in an off-shell formulation of 
eleven-dimensional supergravity (if such a formulation exists). A
similar approach was employed in \cite{bddv} to deduce the existence
of certain auxiliary fields in d=10 {\cal N} = 1 supergrvaity based on
the supersymmetry transformation properties of currents in d=10, 
${\cal N} = 1$ Yang-Mills theory to which the supegravity theory can
couple. A recent discussion of some progress towards an off-shell
formulation of eleven-dimensional supergravity may be found in \cite{cgnn}.

Finally, it would be interesting to compare our description of noncommutative 
open D2-branes in terms of D0-brane degrees of freedom with an alternate 
description directly in terms of D2-brane degrees of freedom (with a B field on 
the D2-brane worldvolume). The relationship between these two descriptions would 
be something like a Seiberg-Witten map (\cite{sw}) for field theories on 
noncommutative spaces with boundaries.

\section*{Acknowledgments}

I would like to thank Micha Berkooz, Michael Douglas, Simeon
Hellerman, Shamit Kachru, Renata  Kallosh, John McGreevy, Mukund
Rangamani, Eva Silverstein, Lenny Susskind, and Wati Taylor for
helpful discussions and comments. I am grateful to the Enrico Fermi
Institute at the  University of Chicago, the Pacific Institute for the
Mathematical Sciences, and  the New High Energy Theory Center at
Rutgers University for hospitality while  parts of this work were
being completed. The work of M.V.R is supported in part by the
Stanford Institute for Theoretical Physics and by NSF grant 9870115.

\appendix

\section{${\cal N} = 1$ in six dimensions}

The actions for $Dp-D(p+4)$ systems and for the Matrix model of
M-theory with M5-branes considered in this paper are written using the
language of $D=6$ supersymmetric field theory with eight supercharges
(${\cal N} = 2$ supersymmetry when reduced to four dimensions).  The
allowed non-gravitational multiplets are vector multiplets,
hypermultiplets and tensor multiplets, however only the first two
arise in the $Dp-D(p+4)$ actions we consider. The most general action
for an arbitrary number of vector and hypermultiplets that is
renormalizable when reduced to four dimensions may be written as 
\beas
{\cal L} &=& {\rm Tr} \left( -{1 \over 4} F_{\mu \nu} F^{\mu \nu} -
{i \over 2} \bar{\lambda}^\rho \gamma^\mu D_\mu \lambda_\rho \right)\\
&&- D_\mu \bar{\Phi}^{\rho} D^{\mu} \Phi_\rho - {i \over 2} \bar{\chi}
\gamma^{\mu} D_\mu \chi\\ 
&&+ i \epsilon^{\alpha \beta} \bar{\chi}
\lambda_\alpha \Phi_\beta - i \epsilon_{\alpha \beta}
\bar{\Phi}^\alpha \bar{\lambda}^\beta \chi \\ 
&& -{1 \over 2} \left( 2
\bar{\Phi}^\alpha {\rm t}^a \Phi_\beta \bar{\Phi}^\beta {\rm t}^a
\Phi_\alpha -  \bar{\Phi}^\alpha {\rm t}^a \Phi_\alpha
\bar{\Phi}^\beta {\rm t}^a \Phi_\beta \right)\\ 
\eeas 
We use
conventions such that $F_{\mu \nu} = \partial_\mu A_\nu  -
\partial_\nu A_\mu + i[A_\mu A_\nu]$ and $D_\mu = \partial_\mu + i
A^a_\mu  t^a$.

The action is invariant under the following supersymmetry
transformations, 
\beas 
\delta A_\mu^a &=& - i \bar{\zeta}^\rho
\gamma_\mu \lambda^a_\rho\\ 
\delta \lambda^a_\rho &=& {1 \over 2} F^a_{\mu \nu} \gamma^{\mu \nu} \zeta_\rho 
+ 2i \zeta_\alpha
\bar{\Phi}^\alpha {\rm t}^a \Phi_\rho - i  \zeta_\rho
\bar{\Phi}^\alpha {\rm t}^a \Phi_\alpha\\ 
\delta \Phi^i_\alpha &=& \epsilon_{\alpha \beta} \bar{\zeta}^\beta \chi^i\\ 
\delta \chi^i &=& 2i \epsilon^{\alpha \beta} D_\mu \Phi^i_\alpha \gamma^\mu 
\zeta_\beta
\eeas

\section{Useful manipulations in six dimensions}

In this paper, we frequently convert ten-dimensional expressions
covariant under  $SO(9,1)$ Lorentz symmetry into six dimensional
expressions for which the  manifest symmetry group is $SO(5,1) \times
SO(4) \sim SO(5,1) \times SU(2)_R  \times SU(2)_L$. To convert bosonic
expressions, the only non-trivial step is to  write various $SO(4)$
representations in terms of the $SU(2) \times SU(2)$  notation.

Firstly, in $SU(2) \times SU(2)$ language, a vector $X^i$ of $SO(4)$
becomes   
\be 
X_{\r \dr} = {1 \over \sqrt{2}} \left(  \ba{cc}  X^6 +
iX^7 & -X^8 +i X^9 \\  
X^8 + i X^9 & X^6 - i X^7  \ea \right) 
\ee
transforming in the $(2,2)$ representation of $SU(2) \times SU(2)$
with the  reality condition 
\[ 
X_{\r \dr} = \e_{\r \s} \e_{\dr \ds}
X^{\s \ds} \; .  
\] 
The normalization is chosen so that 
\[ 
X^i X^i = \bX^{\r \dr} X_{\r \dr} 
\] 
An antisymmetric tensor $A_{ij}$ of $SO(4)$
splits into self-dual and  anti-self-dual parts transforming in the
${\bf 3}$ of $SU(2)_L$ and $SU(2)_R$  respectively. Defining 
\[ 
A^\r {}_\s  = A^{\r \dr} {}_{\s \dr} \; \qquad A^\dr {}_\ds = A^{\r \dr}
{}_{\r \ds} \; .  
\] we have 
\[ 
A^{\r \dr} {}_{\s \ds} = {1 \over 2}
\delta^\r_\s A^\dr {}_\ds + { 1 \over 2} \delta^\dr_\ds A^\r {}_\s  
\]
Alternately, we can describe these tensors as real $SO(3)$ vectors  
\[
A_A = {i \sqrt{2} \over 4} A^{\r \dr} {}_{\s \dr} \s^A_\r {}^\s \;
\qquad A_{\dot{A}} = {i \sqrt{2} \over 4} A^{\r \dr}  {}_{\r \ds}
\s^{\dot{A}}_\dr {}^\ds 
\] where $A$ and $\dot{A}$ are fundamental
$SO(3)_R$ and $SO(3)_L$ indices. With  these normalizations 
\[ 
{1 \over 2} A_{ij} B_{ij} = A_A B_A + A_{\dot{A}} B_{\dot{A}} 
\] 
From a three-index antisymmetric tensor $A_{ijk}$ we define 
\[ 
A_{\r \dr}
\equiv {1 \over 3} \e^{\a \b} \e^{\da \db} A_{\r \da \; \a \db \; \b
\dr} 
\] 
Finally, we will write a four-index antisymmetric tensor
$A_{ijkl}$ as 
\[  
A \equiv {1 \over 24} \e^{ijkl} A_{ijkl} = - {1
\over 12} A^{\r \dr} {}_{\r \ds}  {}^{\s \ds} {}_{\s \dr}  
\] 
To write
ten-dimensional expressions involving fermions in six dimensional
notation we note that a sixteen component Majorana-Weyl spinor in the
${\bf 16}$  of $SO(9,1)$ splits into a pair of spinors in the ${\bf
(4,2,1)}$ and ${\bf  (4,1,2)}$ of $SO(5,1) \times SU(2) \times
SU(2)$. Fermion bilinears may be  reduced using  
\beas 
\bl \gamma^{\mu} \chi  &=& \bl^\r \gamma^{\mu} \c_\r + \bl^{\dr}
\gamma^{\mu}  \c_\dr\\ 
\bl \gamma^{\mu \nu \lambda} \chi  &=& \bl^\r
\gamma^{\mu \nu \lambda} \c_\r +  \bl^{\dr} \gamma^{\mu} \c_\dr\\ 
A_i \bl \gamma^i \chi &=& - \sqrt{2} i \bar{A}^{\r \dr}  \{ \e_{\s \r}
\bl^\s  \chi_\dr + \e^{\dr \ds} \l^\ds \c_r \}\\ 
A_{\mu \nu i} \bl \gamma^{\mu \nu i} \chi &=& - \sqrt{2} i \bar{A}_{\mu \nu}  
{}^{\r
\dr} \{ \e_{\s \r} \bl^\s \gamma^{\mu \nu} \chi_\dr + \e^{\dr \ds}
\l^\ds  \gamma^{\mu \nu} \c_r \}\\ 
{1 \over 2} A_{\mu ij} \bl
\gamma^{\mu ij} \chi  &=& \sqrt{2} i \{  A_{\mu A} \bl^\r \sigma^A_\r
{}^\s \gamma^\mu \chi_\s +  A_{\mu \dot{A}} \bl^\dr
\sigma^{\dot{A}}_\dr {}^\ds \gamma^\mu \chi_\ds \}\\ 
\eeas 
Additional
formulae useful in manipulating expressions in our notation are 
\[ 
\bl \gamma_{\mu_1 \cdots \mu_n} \chi = (-1)^n \bc^c \gamma_{\mu_n \cdots
\mu_1}  \lambda^c 
\] 
\[ 
A_{[\r} B_{\s} C_{\tau]} = 0  
\] 
\[ 
\e_{\r \s}
A_\tau + \e_{\s \tau} A_\r + \e_{\tau \r} A_\s = 0 
\] 
\[ 
\sigma^A_\r
{}^\s \sigma^A_\a {}^\b = 2 \delta_\r^\b \delta_\s^\a - \delta_\r^\s
\delta_\a {}^\b  
\]
Finally, in manipulating expressions with three or
more spinors, it is often  necessary to use Fierz identities which may
be derived from the completeness  relation 
\beas 
M_{\a \b} &=& {1
\over 4} \tr( M \gamma_0) (\gamma^0)_{\a \b} + {1 \over 4} \tr(  M
\gamma_\mu \gamma_0) (\gamma^0 \gamma^\mu)_{\a \b} + {1 \over 8} \tr(
M  \gamma_{\nu \mu} \gamma_0) (\gamma^0 \gamma^{\mu \nu})_{\a \b}\\ 
&& + {1 \over 48}  \tr( M \gamma_{\lambda \nu \mu} \gamma_0) (\gamma^0
\gamma^{\mu \nu \l})_{\a \b} 
\eeas 
For spinors $\lambda$ and $\theta$
of opposite chirality, we find 
\beas 
\theta \bl &=& - {1 \over 4} (\bl
\t)  \identity  - {1 \over 4} (\bl \gamma^{0a}  \t) \gamma^{0a} + { 1
\over 8} (\bl \gamma^{ab} \t) \gamma^{ab} \; .  
\eeas 
For spinors
$\theta$ and $\chi$ of the same chirality, we have 
\[ 
\t \bc = {1
\over 4} (\bc \gamma^{0} \t) \gamma^0 - {1 \over 4} (\bc \gamma^{a}
\t) \gamma^a - {1 \over 8} (\bc \gamma^{0ab} \t) \gamma^{0ab} \; .  
\]

\section{Conserved currents in the Berkooz-Douglas model}

The eleven-dimensional supersymmetry algebra is given by 
\be 
\{Q_\a,
Q_\b \} = 2 P_\mu (\Gamma^\mu \Gamma^0)_{\a \b} + c_2 Z_{\mu \nu}
(\Gamma^{\mu \nu} \Gamma^0)_{\a \b} + c_5 Z_{\mu_1 \cdots \mu_5}
(\Gamma^{\mu_1  \cdots \mu_5} \Gamma^0)_{\a \b}
\label{qalg} 
\ee 
Here
$Z_{\mu \nu}$ and $Z_{\mu_1 \cdots \mu_5}$ are central charges
corresponding to the membrane and five-brane respectively.  In the
presence of a five-brane oriented along the $0,1,2,3,4,5$ directions,
the  preserved supersymmetries are those given by ${\cal P}_+ Q^+
\equiv {1 \over  2}(1 + \Gamma)Q = Q$, where $\Gamma =
\Gamma^{012345}$.

From (\ref{qalg}), the commutator of the preserved supersymmetries
becomes 
\beas 
\{Q^+_\a, Q^+_\b \} &=& 2 P_a ({\cal P}_+ \Gamma^a
\Gamma^0)_{\a \b}\\ 
&& + 2 c_2 Z_{i a} ({\cal P}_+ \Gamma^{ia}
\Gamma^0)_{\a \b}\\ 
&& + 5 c_5 Z_{ijkla} ({\cal P}_+ \Gamma^{ijkla}
\Gamma^0)_{\a \b}\\ 
&& + 10 c_5 Z_{ijabc} ({\cal P}_+ \Gamma^{ijabc}
\Gamma^0)_{\a \b}\\ 
&& +  c_5 Z_{abcde} ({\cal P}_+ \Gamma^{abcde}
\Gamma^0)_{\a \b} 
\eeas 
The central charges appearing on the right
hand side are the momentum in the  5-brane directions, the charge of a
membrane sharing one direction with the  5-brane, and the charge of a
5-brane sharing 1, 3, or 5 directions with the  brane. These central
charges will commute with the Hamiltonian and therefore  correspond to
the conserved currents of the theory. In terms of the momentum  space
currents of the Berkooz-Douglas model, we therefore expect the
following  conservation laws (for further discussion, see \cite{vr})
\beas 
\dot{T}^{++} &=& i k_a T^{+a} + \bar{k}^{\r \dr} T^{+} {}_{\r
\dr}\\ 
\dot{S}^+_{+ \dr} &=& i k_a (S_+^a)_\dr + i \bar{k}^{s \ds}
(S_{+ \s \ds})_\dr\\ 
\dot{T}^+ {}_{\r \dr} &=& i k_a T^a {}_{\r \dr} +
i \bar{k}^{\s \ds} T_{\s  \ds \; \r \dr}\\ 
\dot{J}^{+a} {}_{\r \dr}
&=& i k_b J^{ba} {}_{\r \dr} - {\sqrt{2} \over 2}   k_{\s \dr} \s^A_\r
{}^\s J^{a A} - {\sqrt{2} \over 2}  k_{\r \ds}  \s^{\dot{A}}_\dr
{}^\ds J^{a \dot{A}}\\ 
\dot{S}_{- \r}^+ &=& i k_a (S_-^a)_\r + i
\bar{k}^{s \ds} (S_{- \s \ds})_\r\\ 
\dot{T}^{+-} &=& i k_a T^{a -} + i
\bar{k}^{\s \ds} T^- {}_{\s \ds}\\ 
\dot{J}^{+-a} &=& i k_b J^{b - a} +
i \bar{k}^{\s \ds} J^{- a} {}_{\s \ds}\\ 
\dot{M}^{+-bcde} &=& i k_a
M^{a - bcde} + i \bar{k}^{\s \ds} M_{\s \ds} {}^{- bcde}\\
\dot{M}^{+-A} &=& i k_a M^{a - A} + i \bar{k}^{\s \ds} M_{\s \ds }
{}^{-A}\\ 
\dot{M}^{+-\dot{A}} &=& i k_a M^{a - \dot{A}} + i
\bar{k}^{\s \ds} M_{\s \ds}  {}^{- \dot{A}}\\ 
\dot{M}^{+-} &=& i k_a
M^{a -} + i \bar{k}^{\s \ds} M_{\s \ds} {}^{-}\\ 
\eeas

\bibliographystyle{plain}

\end{document}